\documentclass[onecolumn]{emulateapj}
\usepackage{amsmath,amssymb,graphicx,natbib,subfigure}

\newcommand{\mh}{{M_\bullet}}
\newcommand{\msun}{{M_\odot}}

\newcommand{\beq}{\begin{equation}}
\newcommand{\eeq}{\end{equation}}
\newcommand{\sbh}{SBH}

\begin{document}

\title{Gravitational Encounters and the Evolution of Galactic Nuclei. II. \\ Classical and Resonant Relaxation}
\author{David Merritt}
\affil{Department of Physics and Center for Computational Relativity
and Gravitation, Rochester Institute of Technology, Rochester, NY 14623}

\begin{abstract}
Direct numerical integrations of the Fokker-Planck equation in energy-angular momentum space
are carried out for stars orbiting a supermassive black hole (\sbh) at the center of a galaxy.
The algorithm, which was described in detail in an earlier paper,
includes diffusion coefficients that describe the effects of both random (``classical'')
and correlated (``resonant'') encounters.
Steady-state solutions are similar to the Bahcall-Wolf solution, 
$n(r) \propto r^{-7/4}$,
but are modified at small radii due to the higher rate of diffusion in angular momentum,
which results in a low-density core.
The core radius is a few percent of the influence radius of the \sbh. 
The corresponding phase-space density $f(E,L)$ drops nearly to zero at low energies,
implying almost no stars on tightly-bound orbits about the \sbh.
Steady-state rates of stellar disruption are presented, and a simple analytic expression
is found that reproduces the numerical feeding rates with good accuracy.
The distribution of periapsides of disrupted stars is also computed.
Time-dependent solutions, $f(E,L,t)$, are also computed, starting from initial conditions similar
to those produced by a binary \sbh.
In these models, feeding rates evolve on two timescales: rapid evolution during which
the region evacuated by the massive binary is refilled by angular-momentum diffusion; 
and slower evolution as diffusion in energy causes the density profile at large radii 
to attain the Bahcall-Wolf form.
\end{abstract}


\section{Introduction}
Paper I in this series \citep{Paper1} presented a numerical algorithm for integrating the Fokker-Planck
equation describing $f(E,L,t)$, the phase-space density of stars orbiting  a \sbh\ at the center
of a  galaxy.
The algorithm described in Paper I was similar to that in the pioneering study of \citet{CohnKulsrud1978},
but with a few modifications.
Loss of stars into the \sbh\ was treated more carefully, by adopting a logarithmic grid
in angular momentum and by incorporating a more precise expression for the loss-cone flux.
In addition, the diffusion coefficients in energy, $E$, and angular momentum, $L$,
were allowed to have more general forms than those
derived in the classical theory of Chandrasekhar, H\'enon, Spitzer and others, all of whom assumed random (uncorrelated) interactions, and (with a few notable exceptions, e.g. \citet{Lee1969}) 
ignored relativistic corrections to the equations of motion.

Two characteristic length scales are commonly associated with a supermassive black hole
(\sbh) at the center of a galaxy.
The gravitational radius $r_g$,
\beq
r_g \equiv \frac{G\mh}{c^2} \approx 4.80\times 10^{-8} \left(\frac{\mh}{10^6 \msun}\right) \mathrm{pc} \mathrm{,}
\eeq
is the length scale set by Einstein's equations for a relativistically compact object.
The (gravitational) influence radius, $r_\mathrm{infl}$, has two standard definitions:
either $r_\mathrm{infl}=r_h$, where
\beq
r_h \equiv \frac{G\mh}{\sigma^2} = \left(\frac{c}{\sigma}\right)^2 r_g \approx 0.43 \left(\frac{\mh}{10^6\msun}\right)
\left(\frac{\sigma}{100\; \mathrm{km\ s}^{-1}}\right)^{-2} \mathrm{pc};
\eeq
or $r_\mathrm{infl} = r_m$, defined implicitly via
\beq
M_\star(r<r_m) = 2\mh.
\eeq
The first of these is expressed in terms of $\sigma$, 
the one-dimensional velocity dispersion of stars in the
galactic nucleus, while the second is defined as the radius containing a stellar (distributed) mass
equal to twice $\mh$.
In the Milky Way, $r_h\approx r_m\approx 10^{0.5}$ pc \citep{Schoedel2009,Chatz2015}.

The classical theory of gravitational encounters is valid at distances $r\gtrsim r_m$ from a \sbh.
In this regime, random encounters imply a similar timescale for changes in both $E$ and $L$:
the ``two-body'' relaxation time $T_r$, given by
\begin{eqnarray}\label{Equation:DefineTr}
T_r = \frac{0.34\sigma^3}{G^2 m_\star\rho\ln\Lambda} \approx
1.2\times 10^{10} \!\left(\frac{\sigma}{100\,\mathrm{km\,s}^{-1}}\right)^{\!3} \!\!\left(\frac{\rho}{10^5\,\msun\,\mathrm{pc}^{-3}}\right)^{\!-1} \!\!\left(\frac{m_\star}{\msun}\right)^{\!-1} \!\!\left(\frac{\ln\Lambda}{15}\right)^{\!-1}\!\mathrm{yr}. 
\end{eqnarray}
Here $\rho$ is the stellar mass density,  $m_\star$ is the mass of a single star,
and $\ln\Lambda$ is the Coulomb logarithm \citep{Chandrasekhar1942}.
In Figure~1, the region where equation (\ref{Equation:DefineTr}) 
defines the timescale associated with gravitational encounters is labelled ``Newton.''

If one imagines approaching ever more closely to the \sbh, the unperturbed orbits change in well-defined ways,
implying corresponding changes  in the dominant mode of gravitational interaction between stars.
Starting at a radius of $\sim 10^{-1}r_m$,
\footnote{More precise estimates of radii like this one are presented below.}
orbits are so nearly Keplerian that the assumption
of uncorrelated encounters breaks down. 
This is the regime of ``resonant relaxation'' \citep{RauchTremaine1996} in which changes in $L$
can occur on much shorter timescales than changes in $E$.
The corresponding spatial region is labelled ``Kepler'' in Figure~1.

Still closer to the \sbh, the lowest-order effects of general relativity (GR) begin to make themselves felt.
Orbits experience planar (apsidal) precession due to the 1st post-Newtonian (1PN) corrections to the
 equations of motion.
One consequence is that the coherence time for the resonant interactions described above
is determined by GR in this region.
In addition, the most eccentric orbits at a given energy will precess due to GR at a higher rate
than most other orbits of similar energy, 
causing the former to behave in qualitatively different ways than the latter
in response to perturbations  \citep{MAMW2011}.
The region where gravitational encounters are strongly affected by these 1PN relativistic effects is labelled
``Schwarzschild'' in Figure~1; this region has an outer radius of 
$\sim  10^{-2} r_m$.

\noindent\begin{minipage}{0.60\textwidth}
\includegraphics[width=\linewidth,angle=90.]{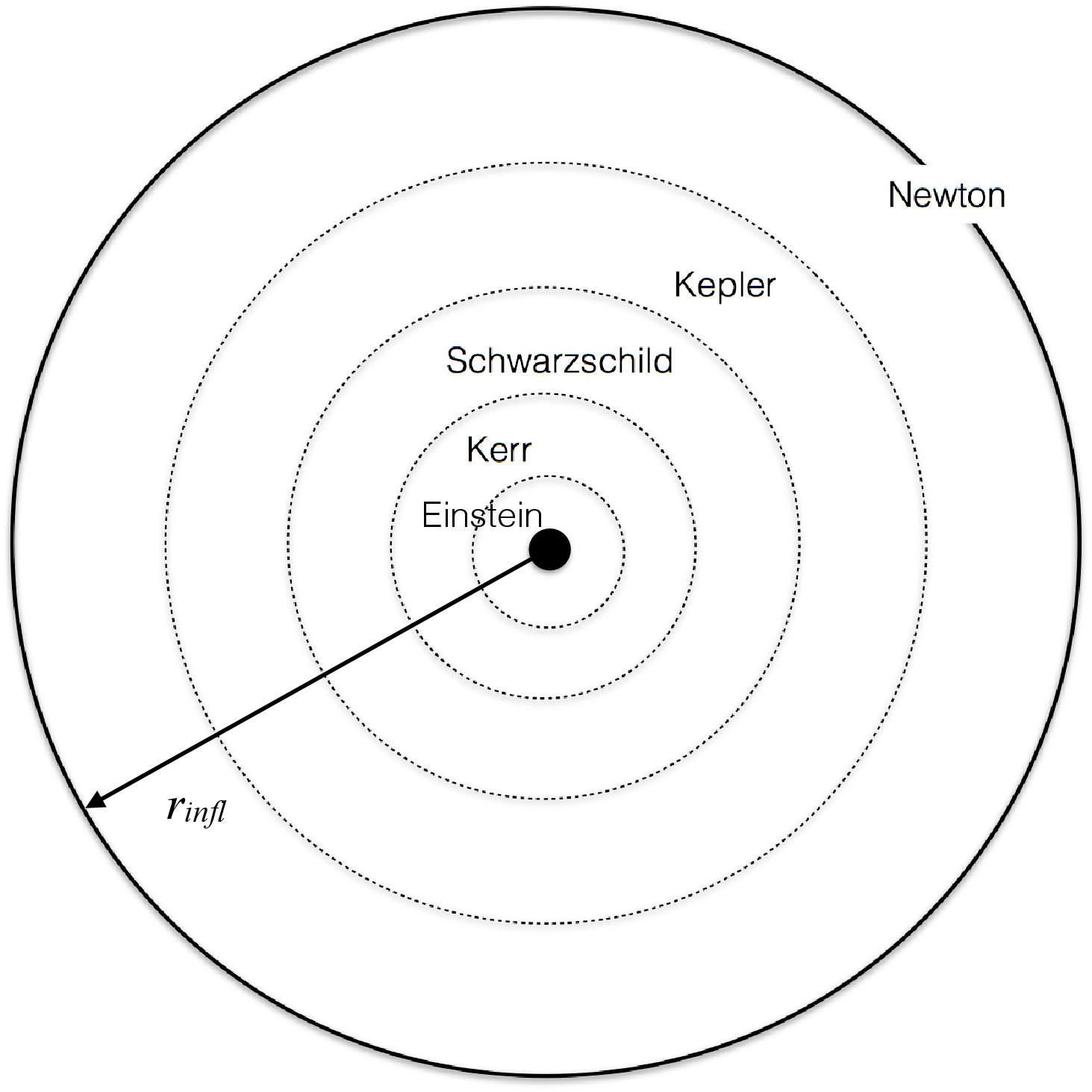}
\end{minipage}
\hfill
\begin{minipage}{0.30\textwidth}
{\footnotesize Fig. 1 -- Sketch of the different regions around a massive black hole at the center of a spherical galaxy.
Each region is defined in terms of the dominant mechanism by which gravitational encounters 
change the orbits of stars.
The outer circle is the black hole's gravitational influence radius $r_\mathrm{infl}$
and the inner circle is its gravitational radius $r_g$.}
\end{minipage}

Another change in the character of the unperturbed motion occurs still closer to the \sbh, where
relativistic frame-dragging causes orbits to precess nodally about the spin axis of the \sbh.
While the importance of frame dragging for gravitational encounters has hardly begun
to be explored, one consequence is  known: within a certain distance, Lense-Thirring torques
from a Kerr black hole  dominate the Newtonian torques that would otherwise  (via ``vector resonant relaxation'') 
be responsible for changes in orbital planes \citep{MAMW2010,MerrittVasiliev2012}.
This region, of outer radius $\sim 10^{-3} r_m$, is labelled ``Kerr'' on Figure~1.

At still smaller radii, PN terms higher than 2nd order can become important, implying
changes in $E$ and $L$ due to gravitational-wave emission.
Within a distance of perhaps $10^2 r_g$ from the \sbh, relativistic corrections 
can not be adequately treated via a Newtonian or post-Newtonian formalism.
This regime is labelled ``Einstein'' in Figure~1.
Of course the number of stars (or compact objects) present at any time in this region is
likely to be  small.

When evaluating the response of stellar orbits near a \sbh\  to gravitational encounters,
most researchers have  applied the classical expressions for the diffusion coefficients,
implicitly assuming that those expressions remain valid arbitrarily close to the \sbh.
An example is the \citet{BahcallWolf1976} steady-state solution for a single-component stellar cluster:
\beq\label{Equation:BWsoln}
f \propto |E|^{1/4},\ \ \ \ n \propto r^{-7/4}
\eeq
which was derived using the \citet{Henon1961} expressions for the orbit-averaged diffusion coefficients
$\langle\Delta E\rangle$, $\langle (\Delta E)^2\rangle$.
This  approach is defensible given that the form taken by the diffusion coefficients in
the regions below ``Newton'' in Figure~1 can not yet be derived from first principles.

However, recent numerical work \citep{Hamers2014,Paper1}
has yielded approximate and fairly general expressions for the angular momentum
diffusion coefficients in the ``Kepler'' regime.  
Those expressions can be included in a Fokker-Planck description and used to evolve $f(E,L)$,
yielding solutions that are valid---if not at all radii---at least to distances $\sim$ten times closer to the \sbh\ than classical solutions like those of \citet{BahcallWolf1976} and \citet{CohnKulsrud1978}.
The diffusion coefficients adopted in the present paper are in fact valid even into the ``Schwarzschild''
regime of Figure~1, in the sense that they correctly account for the effects of 1PN apsidal precession on the coherence time.
However, no attempt is made here to treat the effects of ``anomalous relaxation,'' the qualitatively different
way in which low-$L$ orbits evolve in the Schwarzschild regime \citep{MAMW2011,Hamers2014}.

Section~\ref{Section:Equations} reviews the numerical algorithm used here; 
further details are given in Paper I.
In \S~\ref{Section:BWcusp}, timescales associated with gravitational encounters in the ``Newton,''
``Kepler'' and ``Schwarzschild'' regimes are derived for the case of stars in a Bahcall-Wolf cusp around
 a \sbh.
Section~\ref{Section:Results} presents steady-state and time-dependent solutions for $f(E,L)$.
Section~\ref{Section:Discussion} discusses some implications of the results obtained here for
real stellar systems and \S \ref{Section:Summary} sums up.

\section{Evolution equations}
\label{Section:Equations}
The numerical algorithm for integrating the orbit-averaged Fokker-Planck equation
is described in detail in Paper I.
Features of the algorithm that are most relevant to the current
study are reviewed here.

Stars are assumed to have a single mass, $m_\star$, and to be close enough to the black hole 
(SBH) that the gravitational potential defining their unperturbed orbits is
\beq
\Phi(r) = -\frac{G\mh}{r} \equiv -\psi(r) 
\eeq
with $\mh$ the \sbh\ mass, assumed constant in time.
Unperturbed orbits respect the two isolating integrals $E$, the energy per unit mass,
and $L$, the angular momentum per unit mass.
Following \citet{CohnKulsrud1978} these are replaced by ${\cal E}$ and ${\cal R}$ where
\begin{eqnarray}
{\cal E} \equiv -E = -\frac{v^2}{2} +\psi(r), \ \ 
{\cal R} \equiv \frac{L^2}{L_c^2} ;
\end{eqnarray}
$L_c({\cal E})$ is the angular momentum of a circular orbit of energy ${\cal E}$
so that  $0\le {\cal R}\le 1$.
${\cal E}$ and ${\cal R}$ are related to the semimajor axis $a$ and eccentricity $e$ 
of the Kepler orbit via
\beq\label{Equation:semivsE}
a = \frac{G\mh}{2{\cal E}},\ \ \ \ e^2 = 1 - {\cal R} .
\eeq
The orbital (Kepler) period is
\beq\label{Equation:KeplerPeriod}
P = \frac{2\pi a^{3/2}}{\sqrt{G\mh}}
 = \frac{\pi}{\sqrt{2}} \frac{G\mh}{{\cal E}^{3/2}}
\eeq
and $L_c = G\mh/\sqrt{2{\cal E}}= \sqrt{G\mh a}$.
Spin of the \sbh\ is ignored.

The time dependence of the phase-space number density of stars,
$f({\cal E},{\cal R})$, is described by the orbit-averaged Fokker-Planck equation
\begin{eqnarray}
{\cal J}\frac{\partial f}{\partial t} &=& 
-\frac{\partial}{\partial{\cal E}}\left({\cal J}\phi_{\cal E}\right) 
- {\cal J}\frac{\partial}{\partial {\cal R}}\phi_{\cal R}, \nonumber \\
-\phi_{\cal E} &=& D_{\cal E\cal E}\frac{\partial f}{\partial {\cal E}} + 
D_{\cal E\cal R} \frac{\partial f}{\partial {\cal R}} +
D_{\cal E} f,\ \ 
-\phi_{\cal R} = D_{\cal R\cal E}\frac{\partial f}{\partial {\cal E}} + 
D_{\cal R\cal R} \frac{\partial f}{\partial {\cal R}} + D_{\cal R} f
\label{Equation:FPFluxConserve}
\end{eqnarray}
with flux coefficients
\begin{eqnarray}\label{Equation:DefineFluxCoefs}
D_{\cal E} &=& -\langle\Delta{\cal E}\rangle 
- \frac{5}{4{\cal E}}\langle\left(\Delta{\cal E}\right)^2\rangle
+ \frac12\frac{\partial}{\partial{\cal E}}\langle\left(\Delta{\cal E}\right)^2\rangle
+ \frac12\frac{\partial}{\partial{\cal R}}\langle\Delta{\cal E}\Delta {\cal R}\rangle \;,
\nonumber \\
D_{\cal R} &=& -\langle\Delta{\cal R}\rangle
- \frac{5}{4{\cal E}}\langle\Delta{\cal E}\Delta{\cal R}\rangle 
+ \frac12\frac{\partial}{\partial{\cal E}}\langle\Delta{\cal E}\Delta{\cal R}\rangle 
+ \frac12\frac{\partial}{\partial{\cal R}}\langle\left(\Delta {\cal R}\right)^2\rangle \;,
\nonumber \\
D_{\cal E\cal E} &=& \frac12 \langle\left(\Delta{\cal E}\right)^2\rangle \;,
D_{\cal E\cal R} = D_{\cal R\cal E} = \frac12\langle\Delta{\cal E}\Delta{\cal R}\rangle\; ,
D_{\cal R\cal R} = \frac12\langle\left(\Delta{\cal R}\right)^2\rangle 
\end{eqnarray}
and ${\cal J}\equiv \sqrt{2}\pi^3G^3\mh^3{\cal E}^{-5/2}$
\citep[][5.5.1]{DEGN}.
Quantities in $\langle \; \rangle$ are orbit-averaged diffusion coefficients,
which are expressed as
\begin{eqnarray}\label{Equation:CombinedDiffCoefs}
\langle\Delta{\cal E}\rangle &=&  \langle\Delta{\cal E}\rangle_\mathrm{CK},\ \ \ \ 
\langle\left(\Delta{\cal E}\right)^2\rangle = \langle\left(\Delta{\cal E}\right)^2\rangle_\mathrm{CK}, 
\ \ \ \ 
\langle\Delta{\cal E}\Delta{\cal R}\rangle =  
\langle\Delta{\cal E}\Delta{\cal R}\rangle_\mathrm{CK} ,
\nonumber \\
\langle\Delta{\cal R}\rangle &=& \langle\Delta{\cal R}\rangle_\mathrm{CK}
+ \langle\Delta{\cal R}\rangle_\mathrm{RR}, \ \ \ \ 
\langle\left(\Delta{\cal R}\right)^2\rangle =
\langle\left(\Delta{\cal R}\right)^2\rangle_\mathrm{CK} + 
 \langle\left(\Delta{\cal R}\right)^2\rangle_\mathrm{RR} \;.
\end{eqnarray}
The subscript CK indicates that the diffusion coefficient is computed as in \citet{CohnKulsrud1978};
their derivation was based on standard assumptions about randomness of encounters
\citep{Rosenbluth1957}.
The subscript RR refers to ``resonant relaxation'' \citep{RauchTremaine1996}.
The resonant diffusion coefficients are assumed to have the simple, separable forms
\begin{eqnarray}
\langle \Delta{\cal R}\rangle_\mathrm{RR} = 2 A({\cal E}) \left(1 - 2{\cal R}\right), \ \ \ \ 
\langle\left(\Delta {\cal R}\right)^2\rangle_\mathrm{RR} = 4 A({\cal E}) {\cal R}\left(1-{\cal R}\right) .
\label{Equation:RRDiffCoef}
\end{eqnarray}
The term containing the ${\cal E}$ dependence is
\begin{eqnarray}
A(a) =  \alpha_s^2\left[\frac{M_\star}{\mh}\right]^2 \frac{1}{N} 
\frac{t_\mathrm{coh}}{P^2} , \ \ \ \ \alpha_s = 1.6, \ \ \ \ a = \frac{G\mh}{2{\cal E}} .
\label{Equation:RRDiffCoef2}
\end{eqnarray}
Here $N\equiv N(r<a)$ is the number of stars instantaneously at radii {\bf smaller} than $a$,
$M_\star = m_\star N$, $P$ is the Kepler (radial) period, and $t_\mathrm{coh}$ 
is the coherence time, defined as
\begin{eqnarray}\label{Equation:Definetcoh}
t_\mathrm{coh}^{-1} &\equiv& t_\mathrm{coh,M}^{-1} + t_\mathrm{coh,S}^{-1} \nonumber \\
t_\mathrm{coh,M}(a) &=& \frac{\mh}{Nm_\star}P\;, \ \ \ \ 
t_\mathrm{coh,S}(a) = \frac{1}{12} \frac{a}{r_g} P .
\end{eqnarray}
$t_\mathrm{coh,M}$ is the mean precession time for stars of semimajor axis $a$ due to
the distributed mass around the \sbh\ (``mass precession''), and
$t_\mathrm{coh,S}$ is the mean precession time due to the 1PN corrections to the
Newtonian equations of motion (``Schwarzschild precession'').

\setcounter{figure}{1}
\begin{figure}[h!]
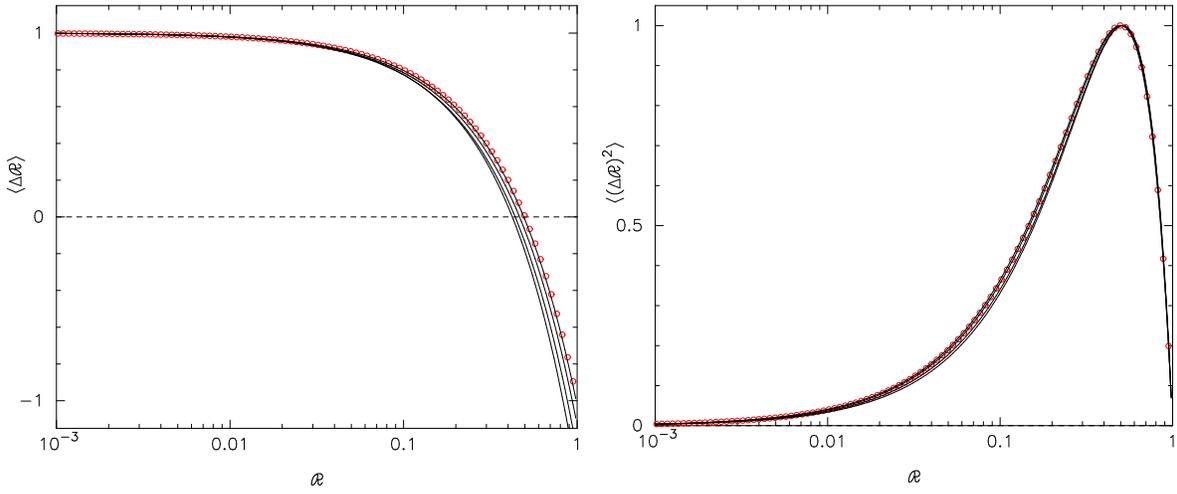

\centering
\mbox{\subfigure{\includegraphics[angle=-90.,width=3.0in]{Figure2A.eps}}\quad
\subfigure{\includegraphics[angle=-90.,width=3.0in]{Figure2B.eps} }}
\caption{Angular momentum diffusion coefficients plotted as functions of ${\cal R}\equiv L^2/L_c^2$.
Open (red) circles show equations~(\ref{Equation:RRDiffCoef}), the adopted expressions for the resonant diffusion coefficients.
Lines show the Cohn-Kulsrud diffusion coefficients, in models having $n(r)\propto r^{-\gamma}$,
i.e. $f\propto {\cal E}^{\gamma-3/2}$, and $\gamma = \{1/2,1,3/2,2\}$.
Curves on the left are normalized to the same $y-$ value at ${\cal R}=0$, and curves on the right are normalized
so as to have the same peak value.
In the left panel, curves with largest $\gamma$ have the smallest ordinate value at
${\cal R}=1$; in the right panel, the curves with $\gamma=0.5$ and $\gamma=2$ have the largest
ordinate values at ${\cal R}\approx 0.05$, while the two curves with $\gamma=1$ and $\gamma=3/2$
lie slightly below.}
 \label{Figure:drr}
\end{figure}

The functional forms chosen for the resonant diffusion coefficients in equations~(\ref{Equation:RRDiffCoef})-(\ref{Equation:Definetcoh})
were shown in Paper I to reproduce the numerically-extracted diffusion coefficients of \citet{Hamers2014},
at least in the particular set of  scale-free nuclear models considered by those authors.
Figure~\ref{Figure:drr} makes another comparison: 
between the ${\cal R}$-dependence of the resonant diffusion coefficients, 
equations (\ref{Equation:RRDiffCoef}),
and the ${\cal R}$-dependence of the classical diffusion coefficients.
The latter were computed from equations (24)-(25) of \citet{CohnKulsrud1978},
assuming scale-free forms for $f$ and $n$:
\beq
f\propto {\cal E}^{\gamma-3/2},\ \ \ \ n \propto r^{-\gamma} 
\eeq
and a $1/r$ potential.
The different curves in Figure~\ref{Figure:drr} were normalized as described in the figure caption;
of course the classical and resonant diffusion coefficients can have very different amplitudes.
It is remarkable that the classical diffusion coefficients are very well fit by the same simple
functions of ${\cal R}$ that were adopted for the resonant diffusion coefficients; and furthermore
that there is so little dependence of the former on $\gamma$.
These results may lend an extra degree of confidence to the functional forms assumed here for the
resonant diffusion coefficients.

Loss of stars into the \sbh\ is controlled by the choice of $r_\mathrm{lc}$, the radius of the physical
loss sphere around the \sbh, and by the conditions imposed on $f$ at the
loss-cone boundary, ${\cal R} = {\cal R}_\mathrm{lc}({\cal E})$, defined as
\begin{eqnarray}\label{Equation:CaptureCond}
{\cal R}_\mathrm{lc} ({\cal E})&=& 
2\frac{\cal E}{{\cal E}_\mathrm{lc}} \left(1-\frac12\frac{\cal E}{{\cal E}_\mathrm{lc}}\right),\ \ 
{\cal E} \le {\cal E}_\mathrm{lc}, \ \ \ \ 
{\cal E}_\mathrm{lc} \equiv \frac{G\mh}{2r_\mathrm{lc}} \;.
\end{eqnarray}
${\cal R}_\mathrm{lc}$ is the normalized angular momentum of an orbit with (Newtonian)
periapsis at $r_\mathrm{lc}$.
The ${\cal R}$-directed flux of stars across the loss-cone boundary is
\beq\label{Equation:DefineFofE}
F({\cal E}) \; d{\cal E}= -{\cal J}({\cal E}) \phi_{{\cal R}}({\cal R}_\mathrm{lc}) \; d{\cal E} 
\equiv -{\cal J}({\cal E})\; \phi_{{\cal R},\mathrm{lc}}({\cal E}) \; d{\cal E} .
\eeq
Two quantities that play important roles in angular momentum diffusion near the loss-cone boundary 
are ${\cal D}$,
\begin{equation}\label{Equation:calDofElc}
{\cal D}({\cal E}) \equiv
\frac{\langle\left(\Delta{\cal R}\right)^2\rangle_t}{2{\cal R}}\bigg|_{{\cal R}={\cal R}_\mathrm{lc}}
= \frac{D_{\cal R R} ({\cal E}, {\cal R}_\mathrm{lc})}{{\cal R}_\mathrm{lc}}
\end{equation}
and $q_\mathrm{lc}$,
\begin{eqnarray}\label{Equation:Defineqlc}
q_\mathrm{lc}({\cal E}) \equiv  \frac{P({\cal E}) {\cal D}({\cal E})}{{\cal R}_\mathrm{lc}({\cal E})}.
\end{eqnarray}
${\cal D}^{-1}$ is effectively an orbit-averaged, angular momentum relaxation time at energy ${\cal E}$.
The quantity $q_\mathrm {lc}$
measures the change in angular momentum per orbital period, compared with the
size of the loss cone.
The loss-cone boundary conditions adopted in all the integrations presented here were the ``Cohn-Kulsrud boundary conditions'' defined in Paper I.
No attempt is made to solve for $f$ inside the loss cone, 
i.e. at ${\cal R}< {\cal R}_\mathrm{lc}$, since $f$ does not satisfy Jeans's theorem in this region.

Solutions are obtained on a ($N_x \times N_z$) grid in $(X,Z)$, where
\begin{eqnarray}
X &\equiv & \ln R = \ln \left[\frac{L}{L_c({\cal E})}\right]^2  , \nonumber \\
Z &\equiv & \ln\left(1 + \beta {\cal E}^*\right) = \ln \left(1+\beta {\cal E}/c^2\right) .
\end{eqnarray}
Integrations presented here used $N_x=N_z=64$ grid points.

The code adopts units such that
\beq
G=\mh=c=1 
\eeq
allowing the results to be scaled to different masses of the \sbh.
Dimensionless parameters that must be specified before the start of an integration include
$m_\star/\mh$, $\ln\Lambda$ and $\Theta_\mathrm{lc}\equiv r_\mathrm{lc}/r_g$.

It is important to emphasize that all the results presented in this paper assume a single mass for the
stars. One reason for this simplification is the current, poor state of knowledge about the form of the resonant 
diffusion coefficients in systems containing a range of stellar masses. 
The mass dependence of the classical diffusion coefficients is of course known; it implies
a rate of energy loss for massive objects that scales in proportion to their mass, leading to
segregation of the more massive objects toward the galaxy center. In steady-state models of the Milky
Way nucleus that contain a realistic stellar mass function \citep{Freitag2006,HopmanAlexander2006L}, 
mass segregation implies that the total density is dominated by the heaviest stellar remnants,
$\sim 10\msun$ black holes, inside a sphere of radius $\sim$ a few $\times 10^{-3}$ pc. 

\section{Important quantities in a Bahcall-Wolf cusp}
\label{Section:BWcusp}

The steady-state numerical solutions presented later in this paper can be described as 
modifications of the classical Bahcall-Wolf solution, equation~(\ref{Equation:BWsoln}).
Here we evaluate some important quantities associated with angular momentum diffusion 
in a nucleus with 
\beq
\rho(r) \equiv m_\star n(r) \propto r^{-7/4}, \ \ \ \ f({\cal E}) \propto {\cal E}^{1/4},\ \ \ \ \psi(r) = \frac{G\mh}{r} ,
\eeq
a unmodified Bahcall-Wolf (1976) cusp.
Like Bahcall and Wolf, we ignore here the $L-$ dependence that a realistic $f$ would 
necessarily have due to capture by the hole.

Define $r_m$ as the radius containing a mass in stars of $2\mh$.
The number density is
\begin{eqnarray}
n(r) = n_0\left(\frac{r}{r_0}\right)^{-7/4} 
= \frac{5}{8\pi} \frac{M_\bullet}{m_\star} \frac{1}{r_m^3}\left(\frac{r}{r_m}\right)^{-7/4} .
\end{eqnarray}
The number of stars with instantaneous radii less than $r$, or semimajor axes less than $a$,
are given respectively by
\begin{subequations}
\begin{eqnarray}
N_r(<r) &=& 2\; \frac{M_\bullet}{m_\star} \left(\frac{r}{r_m}\right)^{5/4},\\
N_a(<a) &=& \frac{\sqrt{\pi}}{2^{3/4}} \frac{\Gamma(11/4)}{\Gamma(5/4)} 
\frac{\mh}{m_\star} \left(\frac{a}{r_m}\right)^{5/4} 
\approx  1.87\; \frac{\mh}{m_\star} \left(\frac{a}{r_m}\right)^{5/4} .
\end{eqnarray}
\end{subequations}
The phase-space number density is
\begin{eqnarray}\label{Equation:fofEBW}
f({\cal E}) &=& \frac{5}{32} \sqrt{\frac{2}{\pi^5}} \frac{\Gamma(11/4)}{\Gamma(5/4)}
\frac{\mh}{m_\star} \frac{{\cal E}^{1/4}}{\left(G\mh\right)^{7/4} r_m^{5/4}} 
= \frac14 \sqrt{\frac{2}{\pi^3}} \frac{\Gamma(11/4)}{\Gamma(5/4)}
\frac{n_0r_0^{7/4}}{\left(G\mh\right)^{7/4}} {\cal E}^{1/4} \nonumber \\
&\approx&0.022\; \frac{\mh}{m_\star} \frac{{\cal E}^{1/4}}{\left(G\mh\right)^{7/4} r_m^{5/4}} 
\approx 0.11\;\frac{n_0r_0^{7/4}{\cal E}^{1/4}}{\left(G\mh\right)^{7/4}} 
\end{eqnarray}
and the distribution of energies is
\beq
N({\cal E})\; d {\cal E} = 4\pi^2 p({\cal E}) f({\cal E})\; d{\cal E}, \ \ \ \ 
p({\cal E}) = \frac{\sqrt{2}\pi}{4}\left(G\mh\right)^3 {\cal E}^{-5/2} .
\eeq

\begin{figure}[h!]
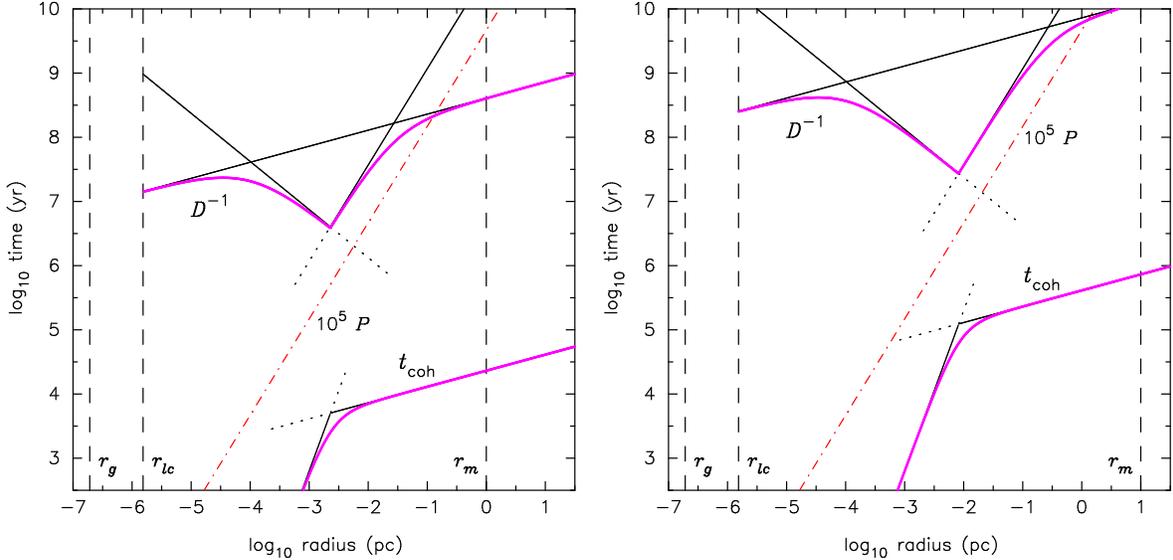

\centering
\mbox{\subfigure{\includegraphics[width=3in]{Figure3A.eps}}\quad
\subfigure{\includegraphics[width=3in]{Figure3B.eps} }}
\caption{Characteristic times in a $n\propto r^{-7/4}$ nucleus.
Parameters are $\mh=4\times 10^6\msun$, $m_\star=1.0\msun$, $r_m=1$ pc
(left) and $r_m=10$ pc (right); the value of $r_m$ at the Galactic center probably
lies between these two values.
The curve labelled $r_\mathrm{lc}$ assumes a capture radius of $8 r_g$, appropriate
for compact objects; tidal disruption of Solar-mass stars would occur at greater distances.
The two solid (black) curves labelled $t_\mathrm{coh}$ show the coherence times due
to mass- and Schwarzschild precession, equations (\ref{Equation:DefinetcohMBW})
and (\ref{Equation:DefinetcohSBW}) respectively;
the mass (Schwarzschild) coherence time has the smaller (larger) value at large radii.
Dot-dashed (red) line is $10^5$ times the Kepler period $P$.
Thick (magenta) curve shows the overall coherence time, equation (\ref{Equation:DefinetcohBW}).
The three curves labelled ${\cal D}^{-1}$  are equations (\ref{Equation:calDNRBW}),
(\ref{Equation:calDRRBWM}), and (\ref{Equation:calDRRBWS}); magenta curve is
equation (\ref{Equation:DefinecalDBW}).
In the case of the curves showing timescales, ``radius'' means ``semimajor axis.''
}
 \label{Figure:times_coh}
\end{figure}

The mass coherence time, equation (\ref{Equation:Definetcoh}), is
\begin{eqnarray}\label{Equation:DefinetcohMBW}
t_\mathrm{coh,M}(a) = \frac{\pi\; r_m^{3/2}}{\sqrt{G\mh}} \left(\frac{a}{r_m}\right)^{1/4} 
\approx 1.2\times 10^5
\left(\frac{\mh}{4\times 10^6\msun}\right)^{-1/2}
\left(\frac{r_m}{3 \mathrm{pc}}\right)^{3/2}
\left(\frac{a}{r_m}\right)^{1/4} \mathrm{yr} 
\end{eqnarray}
and the Schwarzschild coherence time, equation~(\ref{Equation:Definetcoh}b), is
\begin{eqnarray}\label{Equation:DefinetcohSBW}
t_\mathrm{coh,S}(a) = \frac{\pi}{6} \frac{c^2 a^{5/2}}{\left(G\mh\right)^{3/2}} 
\approx  2.0\times 10^{10} \left(\frac{\mh}{4\times 10^6\msun}\right)^{-3/2}
\left(\frac{a}{\mathrm{pc}}\right)^{5/2} \mathrm{yr}.
\end{eqnarray}
These two times are equal when
\begin{equation}\label{Equation:tcohequal}
a = 6^{4/9}\left(r_g^4 r_m^5\right)^{1/9}
\end{equation}
or
\begin{equation}
\frac{a}{r_m} \approx 1.4\times 10^{-3}
 \left(\frac{\mh}{4\times 10^6\msun}\right)^{4/9}
\left(\frac{r_m}{3\mathrm{pc}}\right)^{-4/9} .
\end{equation}
Figure \ref{Figure:times_coh} plots these  times, as well as
the coherence time $t_\mathrm{coh}^{-1} \equiv t_\mathrm{coh,M}^{-1} + t_\mathrm{coh,S}^{-1}$
defined in equation (\ref{Equation:Definetcoh}):
\beq\label{Equation:DefinetcohBW}
t_\mathrm{coh}(a) = \pi\sqrt{\frac{r_m^3}{G\mh}} \left(\frac{a}{r_m}\right)^{1/4}\left(1+\frac{6r_gr_m^{5/4}}{a^{9/4}}\right)^{-1}
\eeq
 assuming $\mh=4\times 10^6\msun$, and for 
two choices of $r_m$: 1 pc and 10 pc, which probably bracket the actual value at the Galactic center
\citep{Schoedel2009,Chatz2015}.

The quantity ${\cal D}$ defined in equation  (\ref{Equation:calDofElc}) 
is effectively an inverse, orbit-averaged, angular momentum relaxation time.
In the case of classical relaxation, the CK diffusion coefficients imply, in the limit ${\cal R}\rightarrow 0$,
\begin{eqnarray}\label{Equation:calDNRBWa}
{\cal D} ({\cal E}) &=& 4\pi\Gamma C f({\cal E}), \ \ \ \ C = \frac{8}{385}\left[-158 + 
45\frac{\Gamma(1/4)}{\Gamma(3/4)}\sqrt{\pi}\right] \approx 1.62
\end{eqnarray}
where $\Gamma\equiv 4\pi (Gm_\star)^2\ln\Lambda$, 
i.e.
\begin{eqnarray}\label{Equation:calDNRBW}
{\cal D}({\cal E})&=& \frac{5C}{\sqrt{2\pi}} \frac{\Gamma(11/4)}{\Gamma(5/4)}
\frac{G^2 m_\star \mh \ln\Lambda}{\left(G\mh\right)^{7/4} r_m^{5/4}} \;{\cal E}^{1/4}
 ,\nonumber \\
{\cal D}^{-1} &\approx& 1.61 \times 10^9 
\left(\frac{\mh}{4\times10^6\msun}\right)^{1/2}
\left(\frac{r_m}{3\ \mathrm{pc}}\right)^{5/4}
\left(\frac{m_\star}{\msun}\right)^{-1}
\left(\frac{\ln\Lambda}{15}\right)^{-1}
\left(\frac{a}{\mathrm{pc}}\right)^{1/4} \mathrm{yr} . 
\end{eqnarray}

In the case that diffusion is dominated by resonant relaxation,
equations~(\ref{Equation:RRDiffCoef}) and (\ref{Equation:RRDiffCoef2}) imply
\beq
{\cal D} = 2A = \frac{2\alpha_s^2}{\pi}
\frac{m_\star}{\mh} \left(\frac{G\mh}{r_m^3}\right)^{1/2}
\frac{t_\mathrm{coh}}{P} \left(\frac{a}{r_m}\right)^{-1/4}.
\eeq
Setting $t_\mathrm{coh} = t_\mathrm{coh,M}$ gives
\begin{eqnarray}\label{Equation:calDRRBWM}
{\cal D} &=& \frac{\alpha_s^2}{\pi}
\frac{m_\star}{\mh} \sqrt{\frac{G\mh}{a^3}}, \nonumber \\
{\cal D}^{-1} &\approx& 3.7\times 10^{10} \left(\frac{\mh/m_\star}{4\times 10^6}\right)
\left(\frac{\mh}{4\times 10^6 \msun}\right)^{-1/2} \left(\frac{a}{\mathrm{pc}}\right)^{3/2} \mathrm{yr}
\end{eqnarray}
while setting $t_\mathrm{coh} = t_\mathrm{coh,S}$ gives
\begin{eqnarray}\label{Equation:calDRRBWS}
{\cal D} &=& \frac{\alpha_s^2}{6\pi}\frac{m_\star}{\mh}
\frac{c^2}{\sqrt{G\mh r_m}} \left(\frac{a}{r_m}\right)^{3/4}, \nonumber \\
{\cal D}^{-1} &\approx& 1.66\times 10^{5} \left(\frac{\mh/m_\star}{4\times 10^6}\right)
\left(\frac{\mh}{4\times 10^6\msun}\right)^{1/2}
\left(\frac{r_m}{\mathrm{3\ pc}}\right)^{5/4}
\left(\frac{a}{\mathrm{pc}}\right)^{-3/4} \mathrm{yr}.
\end{eqnarray}
Equating (\ref{Equation:calDNRBW})
and (\ref{Equation:calDRRBWM})
(which assumes $t_\mathrm{coh} = t_\mathrm{coh,M}$)
yields an estimate of the radius below which resonant relaxation dominates classical relaxation:
\begin{eqnarray}\label{Equation:aRReqNR}
\frac{a_\mathrm{eq}}{r_m} \approx \frac12\left[\frac{4\alpha_s^2}{5\sqrt{\pi}K} \frac{\Gamma(5/4)}{\Gamma(11/4)} 
\frac{1}{\ln\Lambda}
\right]^{4/5} 
\approx 2.8\times 10^{-2} \left(\frac{\ln\Lambda}{15}\right)^{-4/5}.
\end{eqnarray}
This radius can be identified with the sphere labelled ``Kepler'' in Figure~1.

Adopting equation (\ref{Equation:Definetcoh}) for the overall coherence time,
we can write an expression that is valid throughout the resonant-relaxation-dominated regime:
\begin{eqnarray}\label{Equation:DefinecalDBW}
{\cal D} &=& \frac{\alpha_s^2}{\pi}\frac{m_\star}{\mh}
\frac{\sqrt{G\mh}}{a^{3/2}} \left(1+\frac{6r_gr_m^{5/4}}{a^{9/4}}\right)^{-1}, \\
{\cal D}^{-1} &\approx& 3.65\times 10^{10} \left(\frac{\mh/m_\star}{4\times10^6}\right)
\left(\frac{\mh}{4\times 10^6\msun}\right)^{-1/2}
\left(\frac{a}{\mathrm{pc}}\right)^{3/2} \times \nonumber\\
&&\left[1 + 4.55\times 10^{-6} \left(\frac{\mh}{4\times 10^6\msun}\right)\left(\frac{r_m}{3\;\mathrm{pc}}\right)^{5/4}
\left(\frac{a}{\mathrm{pc}}\right)^{-9/4}\right] \mathrm{yr} . \nonumber
\end{eqnarray}
The diffusion time associated with resonant relaxation reaches a minimum when
\beq
a = 3^{4/9}\left(r_g^4r_m^5\right)^{1/9}
\eeq
slightly smaller than the radius at which $t_\mathrm{coh,M} = t_\mathrm{coh,S}$.
Either of these radii can be associated with the sphere labelled ``Schwarzschild'' in
Figure~1.

\section{Results}
\label{Section:Results}

\subsection{Steady-state solutions}
\label{Section:ResultsSS}

\citet{CohnKulsrud1978} obtained various steady-state solutions for $f({\cal E}, {\cal R})$,
assuming classical relaxation, and with parameters chosen to represent stars orbiting a massive black 
hole in a globular cluster.
As they noted, an algorithm that ignores the contribution of the distributed mass to the gravitational
potential can not be expected to correctly represent the solution for $f$ at low binding energies;
that is, beyond the black hole's gravitational influence radius.
In all of their integrations, the outer boundary condition was taken to be
\beq\label{Equation:CKOuterBoundary}
f({\cal E}=0, {\cal R}) = f_0
\eeq
and they identified $f_0$ with $n/(2\pi\langle v^2\rangle)^{3/2}$; $n$ and $\langle v^2\rangle$ 
are respectively the number density and mean-square stellar velocity in the cluster core,
where the density was assumed to be constant with radius.
Cohn \& Kulsrud interpreted their outer boundary condition as describing a fixed,
Maxwellian velocity distribution at large distances from the black hole.

Solutions in this section were computed using  a similar outer boundary condition:
\beq\label{Equation:OuterBoundary}
f^*({\cal E}_\mathrm{min}, {\cal R},t) = f^*({\cal E}_\mathrm{min},{\cal R},0).
\eeq
Here, $f^*$ is the dimensionless phase-space density, and 
${\cal E}_\mathrm{min}$ is the minimum value of ${\cal E}$ on the energy grid.\footnote{It is likely that Cohn \& Kulsrud also enforced their boundary condition at a finite ${\cal E}_\mathrm{min}$, and not at ${\cal E}=0$.}
An approximately equivalent statement is that the outer boundary condition consisted of
specifying a fixed mass density at the outermost grid radius.

Near the loss-cone boundary ${\cal R}={\cal R}_\mathrm{lc}({\cal E})$, 
the Cohn-Kulsrud  conditions were imposed, in the manner described in detail in Paper I.

The initial conditions for $f({\cal E},{\cal R})$ were based on an isotropic power-law model, 
$n\propto r^{-\gamma}$, $f\propto {\cal E}^{\gamma-3/2}$,
 but with a simple modification to account for the presence of the loss cone, namely
 \beq
 f({\cal E}, {\cal R}) = 0,\ \ \ \ {\cal R}\le {\cal R}_\mathrm{lc}({\cal E}).
 \eeq

\begin{figure}[h!]
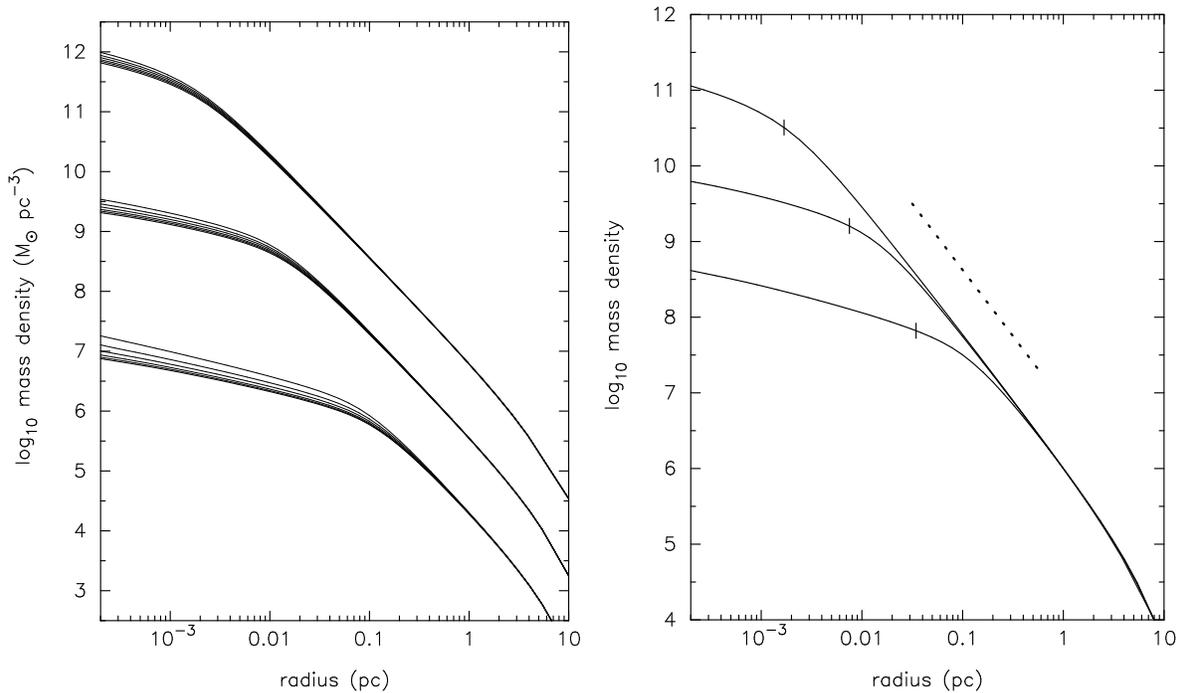

\centering
\mbox{\subfigure{\includegraphics[angle=0.,width=3.0in]{Figure4A.eps}}\quad
\subfigure{\includegraphics[angle=0.,width=3.0in]{Figure4B.eps} }}
\caption{{\bf Left panel:} Steady-state density profiles for integrations with three values of
the outer density normalization and six values of $m_\star/\mh$.
From bottom to top in each set, $m_\star = \{0.3,1,3,10,30,100\}\msun$, assuming
$\mh = 4.0\times 10^6\msun$.
{\bf Right panel:} The three curves from the left panel with $m_\star=1\msun$ have been replotted, 
and rescaled vertically to give the same mass density at $r=1$ pc.
Dotted line has the Bahcall-Wolf slope, $\rho\propto r^{-7/4}$.
Vertical tick marks indicate where $t_\mathrm{coh,M}=t_\mathrm{coh,S}$.
 \label{Figure:rholast}}
\end{figure}

Some examples showing the time-evolution of $n(r,t)$,
starting from initial conditions similar to these, were presented in Paper I.
Here we focus on the steady states.
To minimize integration times, the value of $\gamma$ defining the initial conditions was set
to $7/4$, close to the expected, steady-state value.
Integrations differed in their choice of two parameters: the outer mass density;
and $m_\star/\mh$.
Assuming $\mh=4\times 10^6\msun$ (the code sets the \sbh\ mass to one),
the adopted values of $m_\star$ were
\beq\label{Equation:mstarvalues}
m_\star = \{0.3,1,3,10,30,100\} \msun .
\eeq
In principle, one could identify each value of $m_\star$  with stars of a certain type and estimate the corresponding
tidal disruption radius.
Instead, the radius $r_\mathrm{lc}$ of the loss sphere was chosen to be a fixed multiple of
$r_g$ in in all integrations, $\Theta_\mathrm{lc} \equiv r_\mathrm{lc}/r_g = 15$, i.e.
\beq
r_\mathrm{lc} = 15 r_g \approx 2.9\times 10^{-6} \left(\frac{\mh}{4\times 10^6\msun}\right) \mathrm{pc} ,
\eeq
roughly the value of the tidal-disruption radius for a Solar-type star at the Galactic center.
Many properties of the steady-state solutions, including the rate of loss of stars to the \sbh, are expected to depend only logarithmically on $r_\mathrm{lc}$.

Figure~\ref{Figure:rholast} shows steady-state density profiles for integrations with three different
outer boundary conditions, corresponding to mass densities at one parsec of roughly
\beq
\{1.9\times 10^4, 3.5\times 10^5,   6.1\times 10^6\} \msun \mathrm{pc}^{-3} .
\eeq
These values probably bracket the actual value in the Milky Way \citep{Schoedel2009,Chatz2015}.
The left panel shows solutions for each of the 18 models, i.e., six values of $m_\star$ for
each choice of outer density.
To a good approximation, the form of $\rho(r)$ is determined by the (mass) density normalization,
independent of $m_\star$. 
The reason can be seen by comparing equations (\ref{Equation:calDNRBW}), 
(\ref{Equation:calDRRBWM}) and (\ref{Equation:calDRRBWS}), which 
show that the rate of angular momentum diffusion scales with $m_\star$ in the same way 
for both classical and resonant relaxation, if a fixed value of $r_m$---that is, a fixed mass density---is assumed.
The (weak) dependence of the steady-state density profile on $m_\star$ is due 
to the fact that $q_\mathrm{lc}$, defined in equation (\ref{Equation:Defineqlc}),
is also proportional to $m_\star$.
Solutions with the smallest $m_\star$ approach most closely to the ``empty-loss-cone'' form,
$q_\mathrm{lc}\ll 1$, for which
\begin{eqnarray}\label{Equation:fofERELC}
f({\cal E}, {\cal R}) \approx f({\cal E},1) \frac{\ln({\cal R}/{\cal R}_\mathrm{lc})}{\ln(1/{\cal R}_\mathrm{lc})},
\ \ \  {\cal R}_\mathrm{lc}({\cal E}) \le {\cal R} \le 1
\end{eqnarray}
while large values of $m_\star$ imply $q_\mathrm{lc}\gg 1$ and
\beq\label{Equation:fofERFLC}
f({\cal E}, {\cal R}) \approx \mathrm{const.},\ \ \ {\cal R}_\mathrm{lc}({\cal E}) \le {\cal R} \le 1 ,
\eeq
the ``full-loss-cone'' solution \citep[][6.1.2]{DEGN}.

\begin{figure}[h!]
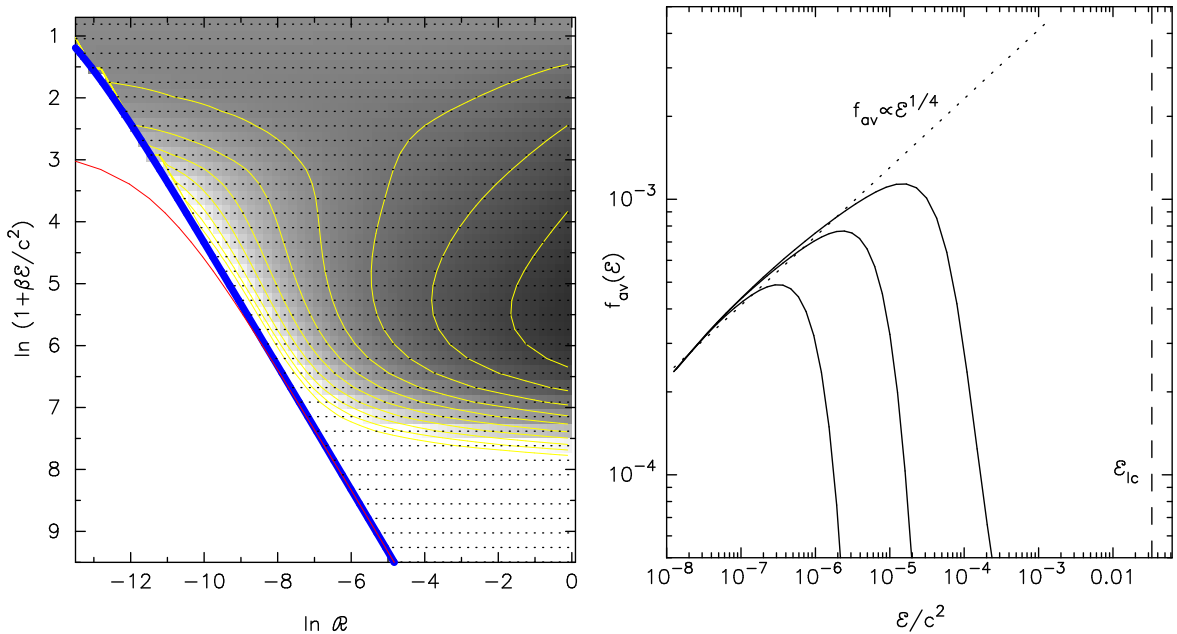

\centering
\mbox{\subfigure{\includegraphics[angle=0.,width=3.in]{Figure5A.eps}}\quad
\subfigure{\includegraphics[angle=0.,width=3.in]{Figure5B.eps} }}
\caption{{\bf Left:} Equilibrium phase-space density in the integration from Figure~\ref{Figure:rholast}
with $m_\star/\msun=1$ and with a final density at $1$ pc of
 $\sim 3.5\times 10^5\msun$ pc$^{-3}$. Greyscale is proportional to $\log f$ and the thin (yellow) curves
 are contours of constant $f$.
 Thick (blue) curve is the loss-cone boundary and thin (red) curve is ${\cal R}_0({\cal E})$.
 The solution grid was uniform in the plotted variables \{$X=\ln{\cal R}$, $Y=\ln(1+\beta{\cal E}/c^2)$\};
 grid centers are indicated with the dots.
{\bf Right:} Angular-momentum-averaged distribution functions for the three, steady-state models from
Figure~\ref{Figure:rholast} with $m_\star = \msun$; the middle curve corresponds to the steady-state
$f$ plotted at left.
The vertical normalization of each curve was chosen to give a fixed value at low energies.
Dotted line shows the Bahcall-Wolf solution and vertical dashed line indicates the energy of a circular
orbit at the assumed radius of the capture sphere around the \sbh.
 \label{Figure:fofelast}}
\end{figure}

Formation of a ``core'' is an expected consequence of resonant relaxation
\citep{HopmanAlexander2006,Madigan2011}.
In the ``Kepler'' regime (Figure~1), diffusion in angular momentum takes place
on a short timescale compared with diffusion in energy.
As a consequence, stars in this region are scattered into the \sbh\ in a time short compared with
the time for the same orbits to be repopulated by (classical) energy diffusion.
An estimate of the value of $a$ below which resonant relaxation dominates classical relaxation
was made in equation~(\ref{Equation:aRReqNR}): $a\approx3\times 10^{-2} r_m$.
Since
\beq
r_m\approx \{0.2, 2.0,  20\} \mathrm{pc} 
\eeq
in the models of Figure~\ref{Figure:rholast}, 
the value of $a$ at transition is predicted to be
$\sim \{6\times 10^{-3},6\times 10^{-2}, 6\times 10^{-1}\}$ pc.
It is reasonable to divide these values by $\sim 2$ to convert from $a$ to radii.
The resulting values are quite similar to the radii of the cores in 
Figure~\ref{Figure:rholast}.

The right panel of Figure~\ref{Figure:rholast} compares the steady-state density profiles
in the three integrations with $m_\star=\msun$.
To assist in the comparison, the curves have been adjusted vertically so as to have the same density
at a radius of one parsec.

\begin{figure}[h!]
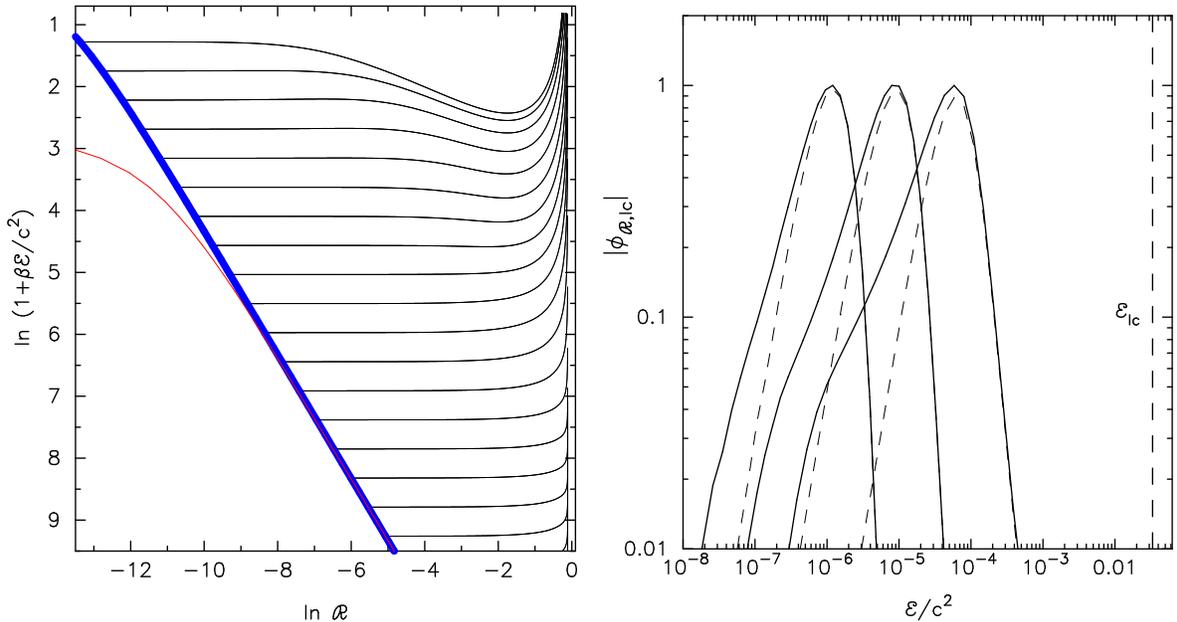

\centering
\mbox{\subfigure{\includegraphics[angle=0.,width=3.in]{Figure6A.eps}}\quad
\subfigure{\includegraphics[angle=0.,width=3.in]{Figure6B.eps} }}
\caption{{\bf Left:} Streamlines of the flux (equation~\ref{Equation:FPFluxConserve}) 
in the steady-state model of Figure~\ref{Figure:fofelast}.
{\bf Right:} Flux of stars into the loss cone as a function of energy.
The three sets of curves correspond to the three models from Figure~\ref{Figure:rholast}.
Each curve has been normalized vertically to give a peak flux of one.
Dashed curves show the contribution to $\phi_{\cal R}$ from resonant relaxation.
\label{Figure:stream}}
\end{figure}

Depletion of $f$ at high binding energies should eventually result in gradients 
with respect to $E$ that drive a (classical) flux that balances the losses due to (resonant) diffusion in $L$.
The left panels of Figures~\ref{Figure:fofelast} and \ref{Figure:stream} provide support for this statement.
Plotted there are the steady-state $f({\cal E}, {\cal R})$ (Figure~\ref{Figure:fofelast}),
and streamlines of the flow in $({\cal E}, {\cal R})$ space (Figure~\ref{Figure:stream}),
of the model from Figure~\ref{Figure:rholast} with $m_\star/\msun=1$ and with 
the intermediate, large-radius density.
There is a remarkably strong depletion of $f$ above a certain binding energy.
The right panel of Figure~\ref{Figure:fofelast} shows angular-momentum-averaged $f$'s:
\beq\label{Equation:DefinefavofE}
\overline{f}({\cal E}) = \int_0^1 f({\cal E},{\cal R})\; d{\cal R} 
\eeq
for the three steady-state models from Figure~\ref{Figure:rholast} with $m_\star/\msun=1$.
To a good approximation, $f=0$ above a certain ${\cal E}$, and so the configuration-space
density at small radii has the form
\beq
n(r) \sim r^{-1/2} ,
\eeq
the density of a population of stars with a single energy moving in a $1/r$ potential.
This is approximately the central dependence of $\rho$ on $r$ in the profiles of Figure~\ref{Figure:rholast}.

The thin (red) curve in the left panels of Figures~\ref{Figure:fofelast} and \ref{Figure:stream} 
is the quantity ${\cal R}_0({\cal E})$,
the $f=0$ intercept of the Cohn-Kulsrud boundary-layer solution extrapolated
inside the loss cone \citep[][equation 6.65]{DEGN}.
An ``empty'' loss cone has ${\cal R}_0\approx {\cal R}_\mathrm{lc}$.
At low binding energies, ${\cal R}_0$ can be seen to drop below ${\cal R}_\mathrm{lc}$,
indicating that the loss cone is becoming progressively fuller far from the \sbh.

The right panel of Figure~\ref{Figure:stream} plots the ${\cal R}$-directed flux,
$\phi_{\cal R}$, at the loss-cone boundary
as a function of energy, in the three steady-state models of Figure~\ref{Figure:rholast}.
Corresponding to the depletion of $f$ at large binding energies, there is a similar depletion
in the loss-cone flux, such that $\phi_{\cal R}$ peaks narrowly around a certain energy.
The dashed curves in this figure show the contribution to the flux from resonant relaxation.
As expected, the resonant contribution to the flux becomes dominant at roughly the same energy where
the depletion in $f$ occurs.

\begin{figure}[h!]
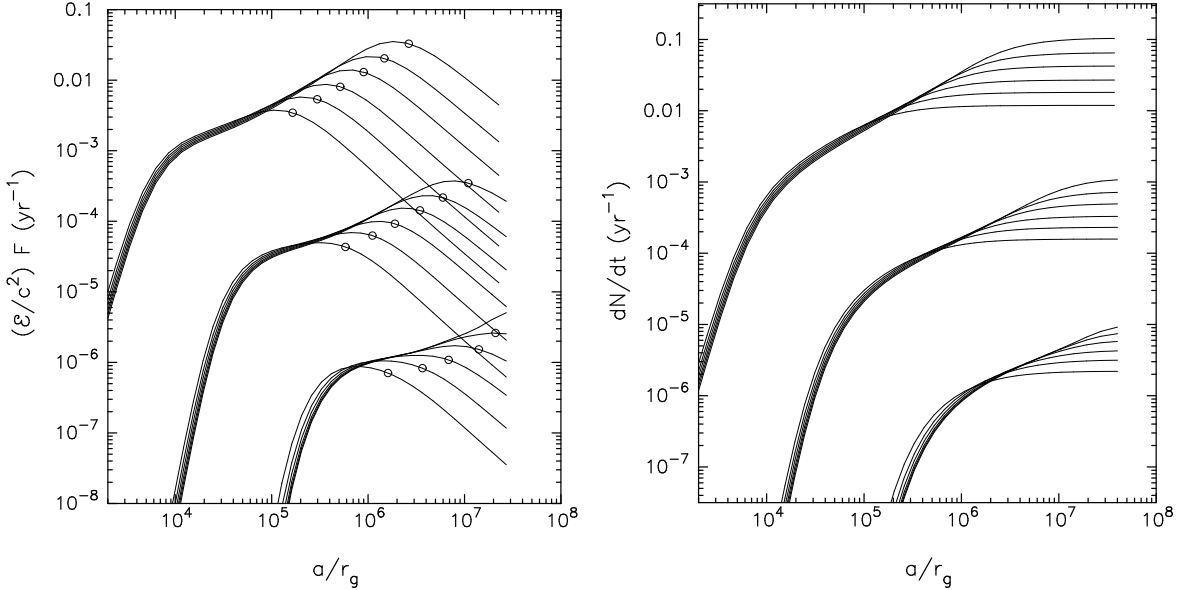

\centering
\mbox{\subfigure{\includegraphics[angle=0.,width=3.in]{Figure7A.eps}}\quad
\subfigure{\includegraphics[angle=0.,width=3.in]{Figure7B.eps} }}
\caption{Local (left) and integrated (right) loss rates, defined as number of stars per year,
 for the equilibrium models of Figure~\ref{Figure:rholast}, assuming $\mh=4\times 10^6\msun$.
In each set of curves, the value of $m_\star$ increases from top to bottom.
The energy ${\cal E}=G\mh/(2a)$
at which $q_\mathrm{lc}=|\ln{\cal R}_\mathrm{lc}|$ is indicated by a circle in the curves on the left.
 \label{Figure:losslast}}
\end{figure}

A quantity more directly related to the loss rate than $\phi_{{\cal R},\mathrm{lc}}({\cal E})$ is 
$F({\cal E}) = -{\cal J}({\cal E}) \phi_{{\cal R},\mathrm{lc}} ({\cal E})$;
equation~(\ref{Equation:DefineFofE}) states that the integral of $F({\cal E})$ with respect to
energy yields $\dot N$.
The left panel of Figure~\ref{Figure:losslast} plots $\left| {\cal E} F({\cal E}) \right| $
 for each of the 18 steady-state models of Figure~\ref{Figure:rholast}.
This quantity can be interpreted as the contribution to the total loss rate from
stars in the energy interval $d{\cal E}/{\cal E}$, or equivalently, $da/a$.
It is clear from this figure that there is a substantial contribution to the feeding rate from stars
at large radii, hence in the classical regime,
particularly in the case of small $m_\star$, i.e. an empty loss cone.
As $m_\star$ is increased (at fixed $\rho$), the radius of transition from full- to empty
loss cones drops; the curves peak,  roughly, at this radius.
In the empty-loss-cone regime, near the \sbh, loss rates (measured in stars per year)
are nearly independent of $m_\star$.
This follows from the dependence ${\cal D}\sim m_\star$, noted above, and the fact
that for a fixed mass density, the number of stars scales as $m_\star^{-1}$.
Far from the \sbh, in the full-loss-cone regime, the expressions derived below imply
\beq
\left|\phi_{{\cal R},\mathrm{lc}}\right| \propto m_\star^{-1}{\cal E}^{11/4},
\ \ \ \ 
\left|{\cal E} {\cal J} \phi_{{\cal R},\mathrm{lc}}\right| \propto m_\star^{-1} {\cal E}^{5/4}, 
\ \ \ \ 
{\cal E} \rightarrow 0.
\eeq
The total loss rate diverges in the case of an everywhere-empty loss cone, $m_\star\rightarrow 0$,
but is finite for finite $m_\star$.

\begin{figure}[h!]
\centering
\includegraphics[angle=0.,width=3.5in]{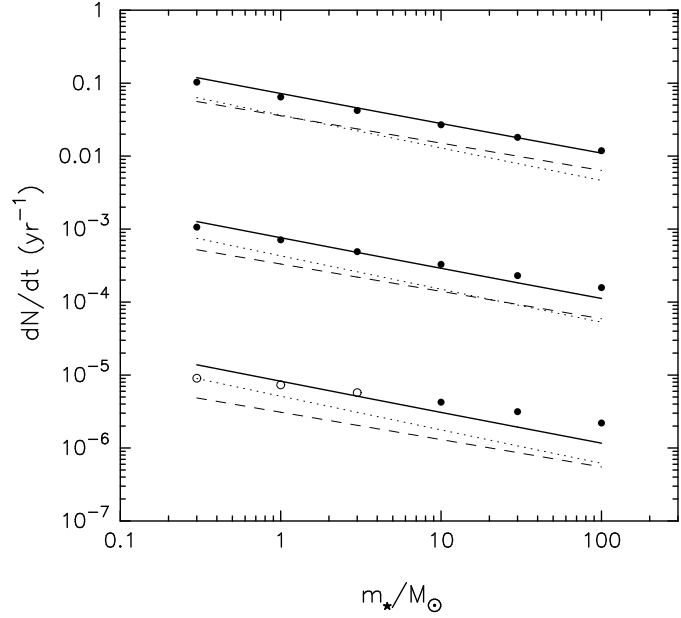}
\caption{Loss rates for the models of Figure~\ref{Figure:losslast}.
Scaling assumes $\mh=4\times 10^6\msun$.   Plotted points are the same numbers given in Table~\ref{Table:ssrates}; dashed and dotted curves
are the approximate analytic relations derived in the text for the full- and empty-loss-cone regimes
respectively, and their sum is shown as the solid curves.
The open circles are from numerical integrations in which the ${\cal E}$- grid probably did not extend
to low enough values to yield the correct, total loss rates (see the right panel of Figure~\ref{Figure:losslast}).
\label{Figure:ssrates}}
\end{figure}

The right panel of Figure~\ref{Figure:losslast} shows integrated loss rates for the same set of models.
As expected, $\dot N$ attains a well-defined limit at low ${\cal E}$, i.e. large $a$, in each model.
These values are listed in Table~\ref{Table:ssrates}, and plotted against $m_\star$ in 
Figure~\ref{Figure:ssrates}.
In three of the models, the adopted energy grid probably did not extend to low enough values of
${\cal E}$ to yield accurate values for the total loss rate; these numbers have been placed in
parentheses in Table~\ref{Table:ssrates}.

It is useful to have an approximate analytic expression for the loss rate.
Since most of the stars lost to the \sbh\ are orbiting in the ``Newton'' regime prior to capture
(Figure~\ref{Figure:losslast}),
it is reasonable to adopt the classical expressions for the angular-momentum 
diffusion coefficients at all energies.
We simplify the derivation even more by (i) adopting for the density profile a Bahcall-Wolf
cusp, unmodified by resonant relaxation and by loss-cone effects; 
and (ii) assuming that stars at any given energy
are either in the empty-loss-cone, or the full-loss-cone, regimes at the time of capture.
In the full-loss-cone (FLC) regime, 
$f$ is assumed to be independent of ${\cal R}$, $f^{\mathrm{FLC}}({\cal E},{\cal R}) = f({\cal E})$.
In the empty-loss-cone (ELC) regime, equation~(\ref{Equation:fofERELC}) for $f$ can be written
\begin{equation}\label{Equation:fofERELCb}
f({\cal R}; {\cal E}) \approx \frac{f({\cal E},1)}{ \ln(1/{\cal R}_\mathrm{lc})}
 \ln\left(\frac{{\cal R}}{{\cal R}_\mathrm{lc}}\right)
\approx  \frac{\overline{f}({\cal E})}{ \ln(1/{\cal R}_\mathrm{lc})+{\cal R}_\mathrm{lc} - 1}
 \ln\left(\frac{{\cal R}}{{\cal R}_\mathrm{lc}}\right),
 \ \ \  {\cal R}_\mathrm{lc} \le {\cal R} \le 1 
\end{equation}
with $\overline{f}({\cal E})$ defined as in equation~(\ref{Equation:DefinefavofE}).
The differential loss rate, $F({\cal E})$, is defined such that the number of stars lost, per unit
of time, from orbits with energies in the range ${\cal E}$ to ${\cal E} + d{\cal E}$ is
$F({\cal E}) d{\cal E}$.
In the FLC regime, the loss rate is equal to the orbital draining rate:
\beq\label{Equation:FofEFLC}
F^{\mathrm{FLC}}({\cal E})  = 4\pi^2 L_c^2({\cal E}) {\cal R}_\mathrm{lc}({\cal E}) f({\cal E})
= P({\cal E})^{-1}  {\cal R}_\mathrm{lc}({\cal E}) N({\cal E})
\eeq
\citep[][equations (6.10,6.72)]{DEGN}.
The loss rate from stars in the ELC regime is
\beq\label{Equation:FofEELC}
F^{\mathrm{ELC}}({\cal E}) = \frac{4\pi^2 L_c^2({\cal E}) P({\cal E}) {\cal D}({\cal E})}
{\ln(1/{\cal R}_\mathrm{lc}) - 1 + {\cal R}_\mathrm{lc}}  \overline{f}({\cal E})
= \frac{{\cal D}({\cal E})  \overline{N}({\cal E})}
{\ln(1/{\cal R}_\mathrm{lc}) - 1 + {\cal R}_\mathrm{lc}}
\approx \frac{{\cal D}({\cal E}) \overline{N}({\cal E})}
{|\ln{\cal R}_\mathrm{lc}|}
\eeq
\citep[][equations (6.59)-(6.62)]{DEGN}, i.e.
\beq\label{Equation:CompareFs}
F^{\mathrm{ELC}}({\cal E}) \approx  
\frac{q_\mathrm{lc}({\cal E})}{\left|\ln {\cal R}_\mathrm{lc}\right|} F^{\mathrm{FLC}}({\cal E}).
\eeq
We assume that equation~(\ref{Equation:FofEFLC}) describes $F({\cal E})$ for 
${\cal E} < {\cal E}_\mathrm{crit}$
and that equation~(\ref{Equation:FofEELC}) describes $F({\cal E})$ for ${\cal E} > {\cal E}_\mathrm{crit}$,
where ${\cal E}_\mathrm{crit}$ is the energy separating the full- and empty-loss-cone regimes.
Identifying $\overline{f}({\cal E})$, $\overline{N}({\cal E})$ and ${\cal D}({\cal E})$ with the expressions for an unmodified Bahcall-Wolf cusp, as given in \S\ref{Section:BWcusp},
the total loss rates from the two regimes can be written in terms of ${\cal E}_\mathrm{crit}$ after integration
over ${\cal E}$, as:
\begin{subequations}\label{Equation:NdotBW}
\begin{eqnarray}
\dot N^{FLC} &\approx& \sqrt{\frac{2}{\pi}} \frac{\Gamma(11/4)}{\Gamma(5/4)} \frac{\mh}{m_\star}
\frac{\left(G\mh\right)^{3/2}}{c^2 r_m^{5/2}} \frac{r_\mathrm{lc}}{r_g} \left(\frac{{\cal E}_\mathrm{crit}}{{\cal E}_m}\right)^{5/4},
\\
\dot N^{ELC} &\approx& \frac{25\sqrt{2}}{32} C\left[\frac{\Gamma(11/4)}{\Gamma(5/4)}\right]^2 
\frac{\ln\Lambda}{\ln{\cal R}_\mathrm{lc}^{-1}}
\sqrt{\frac{G\mh}{r_m^3}} \left(\frac{{\cal E}_\mathrm{crit}}{{\cal E}_m}\right)^{-1},
\ \ \ \ {\cal E}_m\equiv \frac{G\mh}{r_m}.
\end{eqnarray}
\end{subequations}
The constant $C\approx 1.62$ is defined in equation~(\ref{Equation:calDNRBWa}).
In the integral for $\dot N^{ELC}$, the term $\ln{\cal R}_\mathrm{lc}$ was assumed independent of ${\cal E}$;
a reasonable choice for this term might be its value at ${\cal E}={\cal E}_\mathrm{crit}$, i.e.
$\ln{\cal R}({\cal E}_\mathrm{crit})$.

\begin{table}[ht]
\caption{Steady-state loss rates} 
\centering 
\begin{tabular}{lccc} 
\hline\hline 
$\rho\; (r=1\;\mathrm{pc})$ & $m_\star/\msun$ & $\dot N$ (yr$^{-1}$) & $\dot N^\mathrm{ELC}/\dot N$ \\ [0.5ex] 
$(\msun \mathrm{pc}^{-3})$ &  & (numerical) & (analytic) \\ [0.5ex] 
\hline 
$1.9\times 10^4$  & 0.3 & ($9.03\times 10^{-6}$)  & 0.65 \\
                               & 1.0 & $(7.34\times 10^{-6})$  & 0.62\\
                               & 3.0 & $(5.74\times 10^{-6}$)  & 0.60 \\
                               & 10. & $4.24\times 10^{-6}$  & 0.58 \\
                               & 30. & $3.14\times 10^{-6}$  & 0.55 \\
                               & 100. & $2.20\times 10^{-6}$ & 0.53 \\ [1ex] 
$3.5\times 10^5$  & 0.3 & $1.06\times 10^{-3}$  & 0.59 \\
                               & 1.0 & $7.12\times 10^{-4}$  & 0.56 \\
                               & 3.0 & $4.91\times 10^{-4}$  & 0.54 \\
                               & 10. & $3.29\times 10^{-4}$  & 0.52 \\
                               & 30. & $2.31\times 10^{-4}$  & 0.50 \\
                               & 100. & $1.58\times 10^{-4}$  & 0.47 \\ [1ex] 
$6.1\times 10^6$  & 0.3 & $1.03\times 10^{-1}$ &  0.53 \\
                               & 1.0 & $6.45\times 10^{-2}$ &  0.51 \\
                               & 3.0 & $4.22\times 10^{-2}$ &  0.48 \\
                               & 10. & $2.69\times 10^{-2}$ &  0.46 \\
                               & 30. & $1.81\times 10^{-2}$ &  0.44 \\
                               & 100. & $1.18\times 10^{-2}$  & 0.42 \\ [1ex] 
\hline 
\end{tabular}
\label{Table:ssrates} 
\end{table}

We take for ${\cal E}_\mathrm{crit}$ the energy that satisfies
\beq
q_\mathrm{lc}({\cal E}_\mathrm{crit}) = |\ln {\cal R}_\mathrm{lc}| 
\eeq
since at this energy, equation~(\ref{Equation:CompareFs}) suggests that $F^\mathrm{ELC}\approx F^\mathrm{FLC}$.
Using  equations~(\ref{Equation:KeplerPeriod}), (\ref{Equation:Defineqlc}) and (\ref{Equation:calDNRBW})
we find
\begin{eqnarray}
q_\mathrm{lc}({\cal E}) = q_0 \left(\frac{\cal E}{{\cal E}_m}\right)^{-5/4} \frac{m_\star}{\mh} 
\frac{\ln\Lambda}{{\cal R}_\mathrm{lc}({\cal E})}, \ \ \ \ 
q_0 \equiv \frac{5\sqrt{\pi}}{2} \frac{\Gamma(11/4)}{\Gamma(5/4)} C \approx 12.73 .
\nonumber
\end{eqnarray}
When solving for ${\cal E}_\mathrm{crit}\ll {\cal E}_\mathrm{lc}$,
the expression for $q_\mathrm{lc}$ can be simplified by writing ${\cal R}_\mathrm{lc}({\cal E})\approx 2{\cal E}/{\cal E}_\mathrm{lc}$, or
\beq
q({\cal E}_\mathrm{crit}) \approx \frac{q_0}{2} \frac{m_\star}{\mh}\ln\Lambda
\frac{{\cal E}_\mathrm{lc}}{{\cal E}_m}\left(\frac{{\cal E}_\mathrm{crit}}{{\cal E}_m}\right)^{-9/4} .
\eeq
Equating this with $-\ln {\cal R}_\mathrm{lc}$ yields a transcendental equation for 
$x\equiv {\cal E}_\mathrm{crit}/{\cal E}_\mathrm{lc}$:
\beq
x^{9/4} \ln\left(2x\right) =  -\frac{q_0}{2} \frac{m_\star}{\mh}\ln\Lambda \left(\frac{{\cal E}_m}{{\cal E}_\mathrm{lc}}\right)^{5/4} .
\eeq
For values of $x$ in the range of interest ($10^{-7}\lesssim x \lesssim 10^{-5}$),
an approximate solution to $y=x^{9/4} \ln(2x)$ is $x=\sqrt{-2y}$, so that
\begin{subequations}\label{Equation:EcritoverEm}
\begin{eqnarray}\label{Equation:EcritoverEma}
\frac{{\cal E}_\mathrm{crit}}{{\cal E}_m} &\approx& \frac{2.75}{\Theta_\mathrm{lc}^{3/8}}
 \sqrt{\frac{m_\star}{\mh} \ln\Lambda}
\left(\frac{r_m}{r_g}\right)^{3/8} ,\\
\ln{\cal R}_\mathrm{lc}^{-1}  &\approx& -\ln\left[11.0\; \Theta_\mathrm{lc}^{5/8} \sqrt{\frac{m_\star}{\mh}\ln\Lambda}
\left(\frac{r_g}{r_m}\right)^{5/8}
\right] .\label{Equation:EcritoverEmb}
\end{eqnarray}
\end{subequations}
Equations (\ref{Equation:NdotBW}) and (\ref{Equation:EcritoverEm}) are the desired expressions.
The predicted values for $\dot N^\mathrm{FLC}$ and $\dot N^\mathrm{ELC}$
 are plotted as the curves in Figure~\ref{Figure:ssrates}.
The agreement is quite good considering the approximations made;
one implication is that the modifications to the Bahcall-Wolf cusp resulting from resonant relaxation
have little effect on the total loss rate.
The  predicted ratios $\dot N^\mathrm{ELC}/(\dot N^\mathrm{ELC} + \dot N^\mathrm{FLC})$
are given in the final column of Table~\ref{Table:ssrates}; in all cases considered here, 
the two regimes contribute roughly equally to the total loss rate.
We emphasize again that these results apply only to the model considered here, which does not include the 
contribution of the distributed (stellar) mass to the gravitational potential.

\begin{figure}[h!]
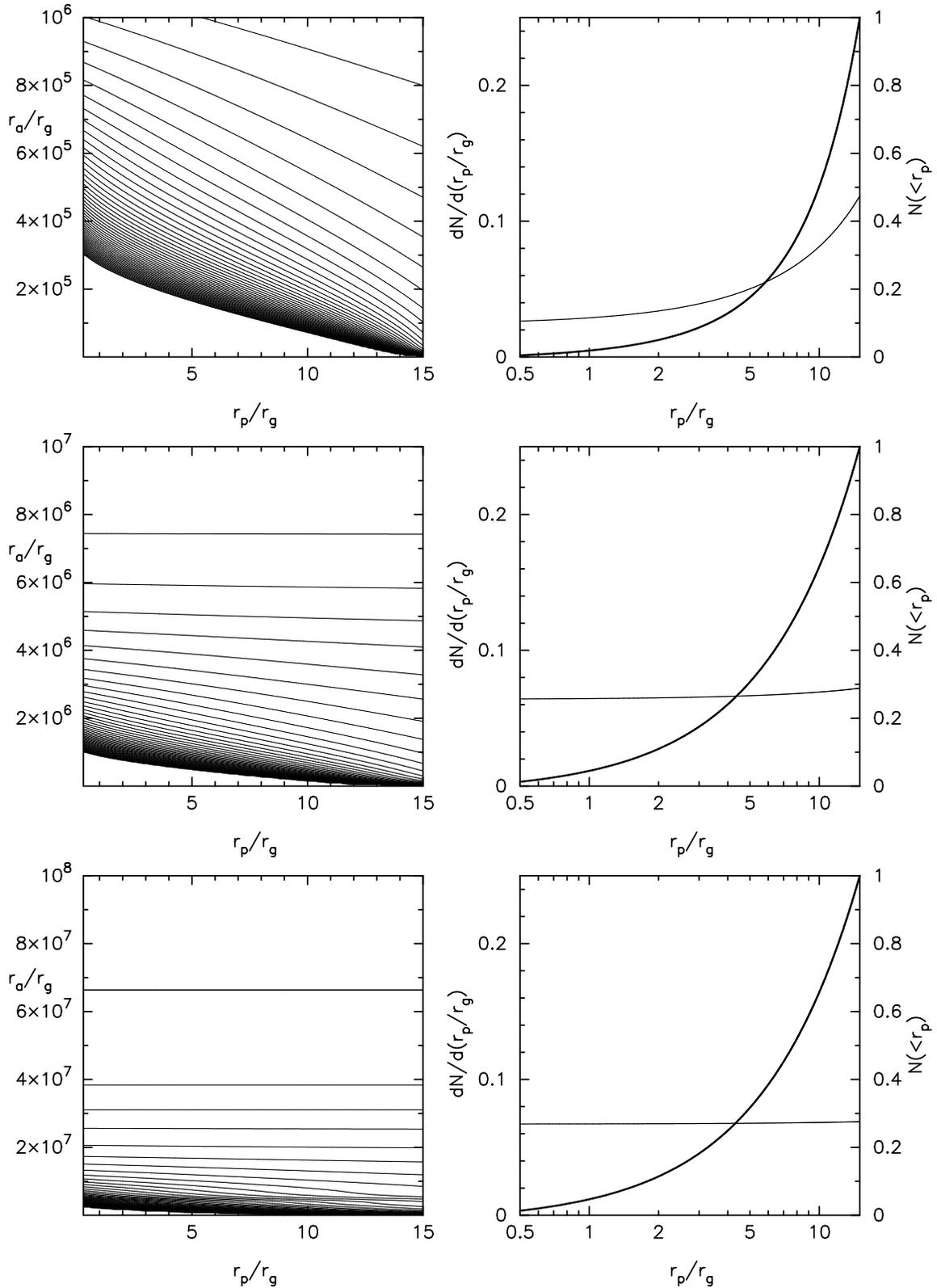

\centering
\includegraphics[angle=-90.,width=6.0in]{Figure9A.eps}
\includegraphics[angle=-90.,width=6.0in]{Figure9B.eps}
\includegraphics[angle=-90.,width=6.0in]{Figure9C.eps}
\caption{Distribution of orbital elements of stars fed to the \sbh, computed as described in
the Appendix, for the three steady-state models from the right-hand panel of Figure~\ref{Figure:rholast}.
{\bf Left:} Joint distribution of $r_a$ and $r_p$, equation~(\ref{Equation:Frarp}). 
Contours are spaced uniformly in $\log F$ and
the range in contour values is $10^3$.
{\bf Right:} Thin curves show $dN/dr_p$, obtained by integrating the function in the left panel 
with respect to $r_a$ at each $r_p$.
Thick curves are the number of stars with periapsides less than $r_p$, obtained by a second
integration with respect to $r_p$. 
Both distributions are normalized to unit total number of stars.
\label{Figure:Frarp}}
\end{figure}

Another interesting property of the steady-state models is the distribution of orbital elements
of the captured stars.
In the empty-loss-cone regime ($q_\mathrm{lc} \ll 1$), loss-cone orbits will have orbital periapsides
close to $r_\mathrm{lc}$, the physical radius of the loss sphere, but in the full-loss-cone regime
($q_\mathrm{lc} \gg 1$), timescales for change in $L$ are comparable with orbital periods
and stars at the time of capture can be on orbits with every value of $r_p$ from zero to $r_\mathrm{lc}$
\citep{CohnKulsrud1978}.
One consequence is that stars can experience stronger tidal stresses than if they were
all lost from orbits with $r_p=r_\mathrm{lc}$ \citep{Gillochon2013}.

Figure~\ref{Figure:Frarp} shows $d^2 F/dr_a dr_p$,
the contribution to the loss-cone flux from stars with 
orbital apoapsides in the range $r_a$ to $r_a +dr_a$ and 
orbital periapsides in the range $r_p$ to $r_p +dr_p$, for three steady-state models.
The details of the calculation are given in the Appendix, which also presents the results of a similar
calculation for the classical Bahcall-Wolf solution.
As shown there, the computed distribution depends on two quantities: $q_\mathrm{lc}({\cal E})$, and
$f({\cal E}, {\cal R}_\mathrm{lc})\equiv f_\mathrm{lc}({\cal E})$.
In the Bahcall-Wolf model, unmodified by resonant relaxation, 
Figure~\ref{Figure:BWrpdist} shows that the distribution of orbital elements is a strong function
of binding energy.
At high ${\cal E}$, i.e. small $r_a$, $q_\mathrm{lc}\ll 1$ and 
the distribution of periapsides is strongly peaked toward $r_p= r_\mathrm{lc}$.
At low ${\cal E}$, i.e. large $r_a$, the loss cone is full and $F$ is nearly independent of
$r_p$.\footnote{These properties of the orbital distribution were noted previously by \citet{Strubbe2011}.}
After integration with respect to $r_a$, the distribution of orbital periapsides in the classical
Bahcall-Wolf solution has the form
shown in the right-hand panel of Figure~\ref{Figure:BWrpdist}, with a maximum at 
$r_p=r_\mathrm{lc}$ and a very steep drop for $r_p\lesssim r_\mathrm{lc}$.

As Figure~\ref{Figure:Frarp} shows, the inclusion of resonant relaxation modifies this distribution,
in the sense of reducing the contribution from stars with $r_p\approx r_\mathrm{lc}$.
This is an indirect consequence of the strong depletion in $f$ at large binding energies.
The depleted orbits are mostly in the empty-loss-cone regime, and their removal implies a 
larger relative contribution to the loss-cone flux from stars in the full-loss-cone regime, hence
$r_p<r_\mathrm{lc}$.
Figure~\ref{Figure:Frarp} shows that this modification is severest in the model with the largest
$r_m$, i.e. the largest core; in this model, the distribution of captured stars
with respect to $r_p$ is nearly uniform.
Even in the model with smallest $r_m$, i.e. the smallest core, the distribution with respect to
$r_p$ is only mildly peaked near $r_\mathrm{lc}$, much less so than in the classical solution.

The distributions shown in Figure~\ref{Figure:Frarp} are computed from models with
$m_\star=\msun$.
Models with larger $m_\star$ have larger $q_\mathrm{lc}$, and the distribution of captured
stars with respect to periapsis in these models is even more uniform than shown in Figure~\ref{Figure:Frarp}.

To a reasonable approximation, therefore, one can write for all these models:
\beq
P(<r_p) \approx \frac{r_p}{r_\mathrm{lc}} ,
\eeq
where $P$ is the probability of capture from an orbit with periapsis less than $r_p$.

\subsection{Time-dependent solutions}

In their pioneering work, Cohn \& Kulsrud (1978) presented only steady-state solutions.
This was in keeping with their parameter choices, which were appropriate to massive black holes
at the centers of globular clusters.
But energy diffusion timescales near the centers of  galaxies are often much longer than in globular 
clusters,
and it is likely that many nuclei have not yet  reached steady states under 
the influence of gravitational encounters.
This is presumably the case in all galaxies with parsec-scale cores; but even in galaxies
with dense, nuclear star clusters, inferred relaxation times are often of order $10^9$ yr or more \citep{Merritt2009}.
The nucleus of the Milky Way probably falls in the non-relaxed category \citep{Merritt2010}.

The time-evolution of such nuclei will differ depending on their assumed initial state.
One widely discussed model invokes a binary \sbh, which scatters and
redistributes stars before (presumably) achieving a small enough separation that
coalescence of the two black holes can occur \citep{BBR1980}.
The late evolution of such binaries is not well understood, but their initial evolution appears
to be fairly robust \citep{MM2001}.
After forming a bound pair, at a separation roughly equal to the influence radius  of the
larger \sbh, the binary separation rapidly decreases through the combined influence
of dynamical friction and three-body interactions with stars.
This phase ends at the so-called ``hard binary'' separation, $a\approx a_h$, where
\beq
a_h \equiv \frac{G\mu}{4\sigma^2} = \frac{M_2}{M_{12}}\frac{r_h}{4}
\eeq
 \citep[][equation 8.23]{DEGN}.
Here, $M_{12}\equiv M_1+M_2$ is the binary mass; $\mu=M_1M_2/M_{12}$;
and $r_h\equiv GM_1/\sigma^2$ is the influence radius of the larger \sbh.
At separations $\lesssim a_h$, 
the binary is able to eject stars with high enough velocities that they escape from the
nucleus.
The binary may ``stall'' at this radius; or, if a mechanism exists for repopulating the depleted orbits,
its semimajor axis can continue to drop.

During the early, rapid phase of its evolution, the binary interacts with stars on orbits having periapses
(defined with respect to the binary center of mass) from $\sim r_h$ to $\sim a_h$ or less.
This interaction modifies orbits with a range of periapses, from $\sim r_h$ down to $\sim a_h$.
Stars on orbits with initial periapses $\lesssim a_h$ are removed entirely from the nucleus.
Here, we approximate  the stellar distribution at the end of this phase simply as
\begin{eqnarray}\label{Equation:fgap}
f(E,L) &=& f_0(E), \ \ \ L\gtrsim L_\mathrm{gap} \nonumber \\
&=& 0, \ \ \ \ \ \ \ \ \ L\lesssim L_\mathrm{gap}, 
\end{eqnarray}
with
\beq\label{Equation:DefineLgap}
L_\mathrm{gap}(E) = Ka_h\sqrt{2\left[E-\Phi(Ka_h)\right]} ,
\ \ K\approx 1 ,
\eeq
the angular momentum of an orbit with periapsis at $Ka_h$.
The corresponding density profile has a core of radius $\sim$ a few $\times  a_h$.

Unless the binary mass ratio is close to unity, the core will be small compared with $r_h$.
In this circumstance, one expects post-binary evolution of the stellar distribution to take place on two
timescales.
Initially, the gap in $f$ at low angular momenta is refilled via diffusion in $L$.
The associated timescale is
\beq
T_\mathrm{gap}\approx \left(\frac{L_\mathrm{gap}}{L_c}\right)^2 {\cal D}^{-1}({\cal E}_\mathrm{gap})
\eeq
with ${\cal E}_\mathrm{gap} \approx G\mh/a_h$.
After a time of $\sim T_\mathrm{gap}$, the phase-space density in the region previously emptied by the
binary will be approximately constant with respect to $L$ at each $E$.

On longer timescales, of order $T_\mathrm{gap} \lesssim \Delta t \lesssim T_r$, 
the distribution of orbital energies will evolve, eventually reaching the Bahcall-Wolf steady state.

Evolution of $f$ in the first phase can be approximated by ignoring energy diffusion
and writing the Fokker-Planck equation at each energy as
\beq\label{Equation:FP1d}
\frac{\partial N}{\partial t} \approx {\cal D} \frac{\partial}{\partial {\cal R}} 
\left({\cal R} \frac{\partial N}{\partial {\cal R}}\right) 
\eeq
\citep{MM2003}.
By changing variables from ${\cal R}$ to $\ell\equiv\sqrt{\cal R}$,
equation~(\ref{Equation:FP1d}) becomes the heat conduction equation in cylindrical coordinates with
radial variable $\ell$ and diffusivity ${\cal D}$ and has a known solution in terms of basis functions
\cite[][6.1.5]{DEGN}.
This solution has been applied to galactic nuclei, assuming for
${\cal D}$ the classical angular-momentum diffusion rate, i.e. ${\cal D}^{-1}\approx T_r$ \citep{MerrittWang2005}.

Initial conditions for the integrations presented here were constructed in the same way as
in the previous section, with the added step of setting $f$ to zero at $L\le L_\mathrm{gap}(E)$.
Equation (\ref{Equation:DefineLgap}) is not quite appropriate here given that $a_h$ is defined
in terms of $\sigma$, the stellar velocity dispersion beyond the \sbh\ influence sphere.
Instead, the maximum periapsis of an evacuated orbit was computed from the roughly equivalent
expression
\beq\label{Equation:Definerpmax}
r_{p,\mathrm{max}} = \frac{\mu}{M_{12}}\frac{r_m}{4} = \frac{q}{(1+q)^2} \frac{r_m}{4}
\eeq
with $q\equiv M_2/M_1$ the binary mass ratio \citep{MerrittSzell2006}.

\begin{table}[ht]
\caption{Evacuated-core models} 
\centering 
\begin{tabular}{lll} 
\hline\hline 
$\gamma$ & $q$ & $r_{p,\mathrm{max}}$ (pc)  \\ [0.5ex] 
\hline 
$3/2$  & $0.01$ & $0.005$ \\
           & $0.03$ & $0.014$ \\
           & $0.1$   & $0.041$ \\ [1ex] 
$2$     & $0.01$ & $0.015$ \\
           & $0.03$ & $0.042$ \\
           & $0.1$   & $0.124$ \\ [1ex] 
\hline
\end{tabular}
\label{Table:evacmodels} 
\end{table}

Six initial models were constructed: one set with $\gamma=3/2$ and one set with $\gamma=2$.
These values are, respectively, less than and greater than the Bahcall-Wolf value $\gamma=7/4$.
For each $\gamma$, the binary mass ratio in equation (\ref{Equation:Definerpmax}) was assigned
one of the three values
\beq
q = \{0.01, 0.03, 0.1\} 
\eeq
and the initial $f$ was set to zero for orbits with periapses less than the 
$r_{p,\mathrm{max}}$ given by equation~(\ref{Equation:Definerpmax}), with $K=1$.
The stellar mass was set to $1\msun$, assuming a \sbh\ mass of $4\times 10^6\msun$,
and $r_\mathrm{lc}$ was set to $15r_g$ as in the previous section.
The other important initial parameter was the value of the mass density at large radii.
This was set to 
\beq
\rho(r=1\;\mathrm{pc}) \approx 4.0\times 10^5 \msun \mathrm{pc}^{-3} \nonumber
\eeq
in the three models with $\gamma=3/2$, and
\beq
\rho(r=1\;\mathrm{pc}) \approx 1.0\times 10^5 \msun \mathrm{pc}^{-3} \nonumber
\eeq
in the models with $\gamma=2$.
The corresponding values of $r_m$ were approximately $2.0$ pc and $6.0$ pc, respectively.
Table~\ref{Table:evacmodels} lists the important parameters for the six initial models.

\begin{figure}[h!]
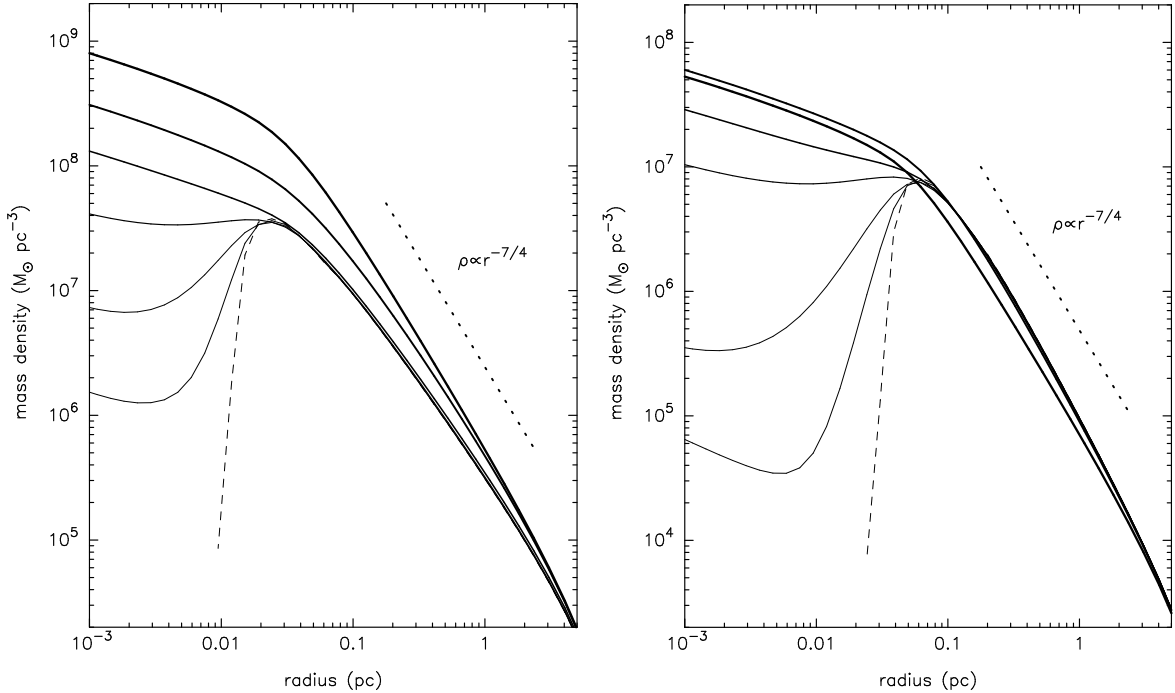

\centering
\mbox{\subfigure{\includegraphics[angle=0.,width=3.in]{Figure10A.eps}}\quad
\subfigure{\includegraphics[angle=0.,width=3.in]{Figure10B.eps} }}
\caption{Evolution of $\rho(r)$ in two integrations starting from evacuated-core initial conditions,
with $q=0.03$.
Left panel: $\gamma=3/2$; right panel: $\gamma=2$.
The initial density profile is shown as the dashed curve;
subsequent times are displayed as curves of increasing width.
Displayed times are 
$\{0.003,0.01,0.03,0.1,0.5,10\}\times 10^9$ yr (left panel) and
$\{0.01,0.03,0.1,0.2,0.5,10\}\times 10^9$ yr (right panel).
Scaling assumes $\mh=4\times 10^6\msun$. 
In the right-hand panel, note that the final central density is slightly lower than its value 
at $t=5\times 10^8$ yr.
\label{Figure:rhooft}}
\end{figure}

Figure~\ref{Figure:rhooft} shows the evolution of the mass density profile in the two
integrations with $q=0.03$.
The values of $r_{p,\mathrm{max}}$ were $\sim 0.014$ pc ($\gamma=3/2$) and
$\sim 0.042$ pc ($\gamma=2$).
The two evolutionary timescales discussed above are evident.
After a time of $\sim 10^8$ yr ($\gamma=3/2$) or $\sim 3\times 10^8$ yr ($\gamma=2$), 
the initial core is erased; during this time, the density at $r\gg r_{p,\mathrm{max}}$ hardly changes.
Over the next $10^9-10^{10}$ yr, energy relaxation causes $\rho(r)$ to approach the
Bahcall-Wolf form at large radii.

Evolutionary models like these are characterized by an additional dimensionless parameter: 
the ratio of the initial core size, $\sim r_{p,\mathrm{max}}$, to the size of the core that forms via
resonant relaxation, in the manner discussed in the previous section.
Adopting the estimate given above, $r_c\approx 0.03r_m$, for the latter core size, 
the ratio becomes
\beq\label{Equation:rcratio}
\frac{r_{p,\mathrm{max}}}{r_c} \approx 8\; \frac{q}{(1+q)^2} .
\eeq
This ratio is $\sim 0.25$ for both of the models of Figure~\ref{Figure:rhooft}, implying that
the final core  should be somewhat larger than the initial core, as seen in the figure.
There is however a change in the structure of the core.
Initially, the core is formed by the exclusion of orbits with small periapses, implying a
zero configuration-space density below some radius.
The core that forms at late times is characterized by a deficit of orbits with high binding
energies; as discussed above, the implied density is nonzero, $\rho \sim r^{-1/2}$ near the center,
due to orbits with low binding energies and small angular momenta that pass near the center.

The fact that the ratio in equation~(\ref{Equation:rcratio}) is less than unity for these models
implies that the initial timescale for core refilling is set by resonant, and not classical,
relaxation.
We can estimate that time from equation~(\ref{Equation:calDRRBWM}),
replacing $a$ by $r_{p,\mathrm{max}}$.
The result is $\sim 6\times 10^7$ yr ($\gamma=3/2$) and $\sim 3\times 10^8$ yr
($\gamma=2$), quite consistent with the evolution seen in Figure~\ref{Figure:rhooft}.

\begin{figure}[h!]
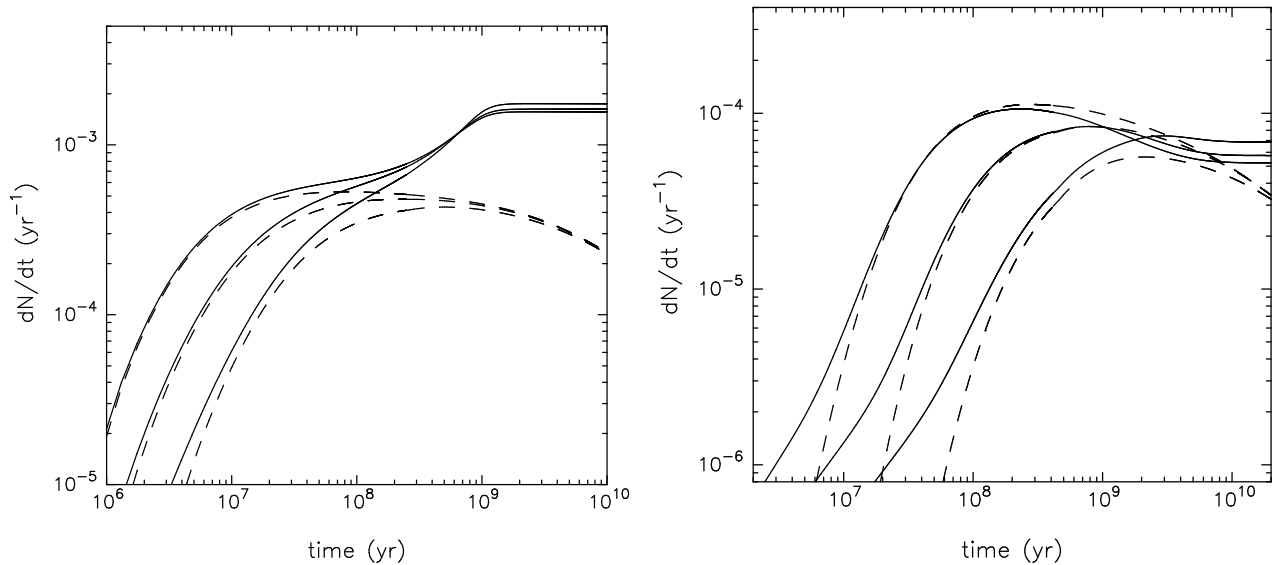

\centering
\mbox{\subfigure{\includegraphics[angle=0.,width=3.25in]{Figure11A.eps}}\quad
\subfigure{\includegraphics[angle=0.,width=3.25in]{Figure11B.eps} }}
\caption{Total loss rates for the six models with evacuated cores at $t=0$.
Left: $\gamma=3/2$; right: $\gamma=2$.
Scaling assumes $\mh=4\times 10^6\msun$.  
Dashed curves are from integrations starting from the same initial conditions, but with the
energy diffusion terms artificially set to zero.
\label{Figure:Ndotoft}}
\end{figure}

The two evolutionary timescales are reflected also in the change with time of the \sbh\ feeding
rate $\dot N$.
Figure~\ref{Figure:Ndotoft} plots this quantity for each of the six models with initially evacuated cores.
Also plotted there, as the dashed curves, are the feeding rates from a second set of integrations
in which the energy diffusion terms were artificially set to zero.
In those models, evolution of $f$ at each $E$ is described approximately by equation~(\ref{Equation:FP1d}).
Comparison of the two sets of curves shows in a very direct way how the early evolution is determined
by angular-momentum diffusion and the late evolution by energy diffusion.

In the models of Figure~\ref{Figure:Ndotoft}, steady-state loss rates are only achieved after
a time of $\sim 10^9$ yr ($\gamma=3/2$) or $\sim 3\times 10^9$ yr ($\gamma=2$).
These times are fixed by the classical relaxation time, equation~(\ref{Equation:DefineTr}),
and hence by the adopted density normalization.
It is clear from this small set of examples that many nuclei, including that of the Milky Way,
might not be in a steady state with regard to \sbh\ feeding rates.
Indeed, in nuclei with \sbh\ masses larger than the value assumed here ($4\times 10^6\msun$),
equation (\ref{Equation:calDRRBWM}) implies that even the initial timescale for core refilling 
due to angular momentum diffusion could exceed $10^{10}$ yr.

\section{Discussion}
\label{Section:Discussion}
\subsection{Comparisons with earlier work}

The consequences of resonant relaxation for the steady-state distribution of stars around a \sbh\ have
been discussed by earlier authors \citep{HopmanAlexander2006,Madigan2011} using more
approximate methods. 
Here we compare the results of those studies with the results obtained here.

\citet{HopmanAlexander2006} derived steady-state solutions for $N(E)$ in the fixed gravitational potential of a \sbh.
Their evolution equation had the form
\beq\label{Equation:fpmodRR}
\frac{\partial N}{\partial t} = -\frac{\partial F_{\cal E}}{\partial {\cal E}} -
F_\mathrm{NR}({\cal E},t) - \chi\; \frac{N}{T_\mathrm{RR}} \;.
\eeq
The term $F_\mathrm{NR}$ accounts for loss of stars into the \sbh\ via ``non-resonant,'' 
i.e. classical, diffusion in $L$; \citet{HopmanAlexander2006} adopted an expression similar to that
used by \citet{BahcallWolf1977} in their earlier study of the 1d problem.
The last term on the right hand side of equation (\ref{Equation:fpmodRR}) 
approximates the loss rate at energy ${\cal E}$ due to resonant relaxation; 
$T_\mathrm{RR}$ is an estimate of the resonant relaxation time,
and the dimensionless factor $\chi={\cal O} (1)$ was included to parametrize uncertainties
in the efficiency of resonant relaxation and in the degree of depletion of phase space near the loss-cone boundary; the latter could not be modeled in their study due to the 1d, $f=f({\cal E})$ approximation.

We can cast our Fokker-Planck equation for $f({\cal E}, {\cal R},t)$ in an analogous form, as follows.
Equation (\ref{Equation:FPFluxConserve}) is
\beq\label{Equation:FPE1}
\frac{\partial N({\cal E}, {\cal R})}{\partial t} = \ldots \; {\cal J} \frac{\partial}{\partial {\cal R}}
\left(D_{\cal RR} \frac{\partial f}{\partial {\cal R}} + D_{\cal R} f \right)
\eeq
where terms depending on the energy-space flux have been omitted.
Adopting the expressions (\ref{Equation:RRDiffCoef}) for the flux coefficients due to RR,
equation (\ref{Equation:FPE1}) becomes
\beq\label{Equation:FPE2}
\frac{\partial N({\cal E}, {\cal R})}{\partial t} = \ldots \; 2A({\cal E}) \frac{\partial}{\partial {\cal R}}
\left[{\cal R}\left(1-{\cal R}\right)\frac{\partial N}{\partial {\cal R}}\right] .
\eeq
Integrating this expression $d{\cal R}$ yields an evolutionary equation for $N({\cal E})$:
\begin{subequations}\label{Equation:FPE3}
\begin{eqnarray}
\frac{\partial N({\cal E}, t)}{\partial t} &=& \ldots \; 2A({\cal E}) 
\left[{\cal R}\left(1-{\cal R}\right)\frac{\partial N}{\partial {\cal R}}\right]_{{\cal R}_\mathrm{lc}}^1
\label{Equation:FPE3a}\\
&=& \ldots -2 A({\cal E}) {\cal R}_\mathrm{lc}({\cal E}) 
\left[1-{\cal R}_\mathrm{lc}({\cal E})\right]\left(\frac{\partial N}{\partial {\cal R}}\right)_{{\cal R}_\mathrm{lc}} .\label{Equation:FPE3b}
\end{eqnarray}
\end{subequations}
To make further progress we need an ansatz for the ${\cal R}$-dependence of $N$.
A natural choice is the ``empty-loss-cone'' solution (\ref{Equation:fofERELCb}),
which implies
\begin{eqnarray}\label{Equation:ELCSoln}
N({\cal R}; {\cal E}) = N(1;{\cal E}) \left(1-\frac{\ln {\cal R}}{\ln {\cal R}_\mathrm{lc}}\right)
= \frac{N({\cal E}) \ln\left({\cal R}/{\cal R}_\mathrm{lc}\right)}{\ln{\cal R}_\mathrm{lc}^{-1}
+ {\cal R}_\mathrm{lc} - 1} 
\end{eqnarray}
where we have identified $N({\cal E}) = \int N({\cal R}; {\cal E}) d{\cal R}$.
Inserting (\ref{Equation:ELCSoln}) into (\ref{Equation:FPE3b}) then yields
\beq
\frac{\partial N}{\partial t} \approx \ldots \; - \frac{2A({\cal E}) N({\cal E})}{\ln (1/{\cal R}_\mathrm{lc})}
\eeq
and comparing this expression with (\ref{Equation:fpmodRR}), we conclude
\beq
\chi \approx 2 \; \frac{T_\mathrm{RR} ({\cal E}) A({\cal E})}{\ln (1/{\cal R}_\mathrm{lc})} .
\eeq
Adopting equation (\ref{Equation:RRDiffCoef2}) for $A({\cal E})$, and \citet{HopmanAlexander2006}'s expression for $T_\mathrm{RR}$:
\beq
T_\mathrm{RR} \equiv \frac{A_\mathrm{RR}}{N_\star(<a)}
\left(\frac{\mh}{m_\star}\right)^2
\frac{P^2(a)}{t_\mathrm{coh}(a)},\ \ \ \ 
A_\mathrm{RR} \approx 3.56
\eeq
then yields
\beq
\chi \approx \frac{18}{\ln{\cal R}_\mathrm{lc}^{-1}} .
\eeq
In the models computed here, equation~(\ref{Equation:EcritoverEmb}) gives 
$\ln {\cal R}_\mathrm{lc}^{-1}\approx 12$, so that $\chi\approx 1.5$.
\citet{HopmanAlexander2006} presented steady-state solutions for the cases
$\chi = \{1,3,10\}$.
Their Figure 8 shows steady-state density profiles, $\rho(r)$, for the Milky Way nucleus,
assuming $\chi = 1$ and $\chi = 10$; there are depletions with respect to the $\chi=0$ (classical) case
inside radii of $\sim 0.05$ pc and $\sim 0.2$ pc, respectively.
While \citet{HopmanAlexander2006} do not clearly state the density normalization for their models
(there are no units on the vertical axis of their Figure 8),
the mass density at $1$ pc appears to be $\sim 1\times 10^5\msun$ pc$^{-3}$, which would place
it midway between the two models in Figure~\ref{Figure:rholast} with the lower density normalizations.
The core sizes in those models are similar to the values found by \citet{HopmanAlexander2006}.

One issue that complicates a comparison of Hopman \& Alexander's (2006) 
results with ours is the choice that they made for the coherence time.
They defined
\beq\label{Equation:TcohHA}
\frac{1}{t_\mathrm{coh}} \equiv
\left|
\frac{1}{t_\mathrm{coh,M}} -
\frac{1}{t_\mathrm{coh,S}}\right| ,
\eeq
with  $t_\mathrm{coh,M}$ and $t_\mathrm{coh,S}$ defined in essentially the same
way as here (equation~\ref{Equation:Definetcoh}).
Both quantities are positive by definition, 
and the minus sign in equation (\ref{Equation:TcohHA}) was said to account for the fact that mass precession is retrograde and Schwarzschild precession prograde.
At a certain energy / radius, the coherence time as defined by equation (\ref{Equation:TcohHA}) becomes infinite.
While the precession rate of a single orbit can be zero
(if its eccentricity has precisely the right value), orbits of other stars at similar radii will still precess,  implying a finite coherence time at every radius.

A very different approach to the problem was taken by \citet{Madigan2011}, 
who developed an ad-hoc statistical model for the effects of resonant relaxation,
calibrated against $N$-body simulations.
Nevertheless, their and our results about the depletion of $f(E)$ at high binding energies and the corresponding flattening of the density profile are qualitatively similar.
The size of the steady-state core predicted by \citet{Madigan2011} for the Milky Way was
$\lesssim 0.05$ pc, consistent with the core sizes in some of the models shown here in Figure~\ref{Figure:rholast}.

\subsection{Comparison with the distribution of stars at the center of the Milky Way}

The distribution of stars near the center of the Milky Way has long been known
to depart from the Bahcall-Wolf steady-state form.
Solutions of the isotropic Fokker-Planck equation that include the stellar potential suggest that 
the Bahcall-Wolf solution (which ignores the stellar potential) should be valid out to distances
of at least $\sim 0.2\; r_m$ from the \sbh\ \citep[][7.1.1]{DEGN}.
Applied to the Milky Way, this result implies that the $n\sim r^{-7/4}$ cusp,
if present, would extend outward to $\sim 0.2-0.6$ pc.
However, number counts of the late-type (i.e. old) stars fail to show a cusp.
Instead, the density of these stars (most of which are believed to be red giants) 
rises only very slowly, if at all, toward the center inside a projected radius of $\sim 0.5$ pc \citep{Buchholz2009,Do2009,Bartko2010}.

\citet{Madigan2011} proposed that resonant relaxation was responsible for the Milky Way core.
However the steady-state cores found by those authors were about an order of
magnitude smaller than the core observed in the Milky Way.
The core sizes in \citet{HopmanAlexander2006}'s steady-state models were also 
substantially smaller than $0.5$ pc, as noted above.
The arguments presented here  (\S~\ref{Section:BWcusp}) suggest that the size of a steady-state
core that is produced by the action of resonant relaxation should scale roughly with $r_m$.
Based on the results plotted in Figure~\ref{Figure:rholast}, a steady-state core as large as $\sim 0.5$ pc in the Milky Way would require $r_m\gtrsim 20$ pc, far larger than most estimates of the \sbh\ influence radius.

A likely resolution to this apparent discrepancy is to assume that the Milky Way nucleus is not
yet in a steady state with regard to diffusion in energy \citep{Merritt2010}.
The observed core could then result from a combination of initial conditions, which have not yet
been erased by the effects of gravitational encounters; and the depleting effects of resonant relaxation.
This possibility is explored in more detail in a separate paper \citep{MAV2015}.

\subsection{Other consequences of the depletion of $f$ at large binding energies}

The striking depletion in $N(E)$ at large binding energies found here, and in some earlier studies,
is a consequence of two things:
(i) the sudden drop in the angular momentum diffusion time below a certain energy,
due to resonant relaxation; and
(ii) the absence of any mechanism that might maintain a high density of stars near the \sbh\
in spite of the high loss rate.
The first assumption seems robust; the second less so, since the forms for the diffusion coeffcients
adopted here are only likely to be valid beyond a certain distance from the \sbh.
If diffusion times become long again in the ``Schwarzschild,'' ``Kerr'' etc. regions of Figure~1,
a high density of stars (or compact objects) might be maintained very near the \sbh.
Ongoing star formation in this region could also maintain a nonzero $f$.

Assuming for the moment that neither happens, one can speculate about the consequences
of the depletion.
An approximate representation of the steady-state phase space density  in 
the models of Figures~\ref{Figure:rholast} and~\ref{Figure:fofelast} is
\begin{eqnarray}\label{Equation:fcore}
f({\cal E},{\cal R}) &\approx& \overline{f}({\cal E}), \ \ \ {\cal E}\lesssim {\cal E}_\mathrm{eq} \nonumber \\
&\approx& 0, \ \ \ \ \ \ \ {\cal E}\gtrsim {\cal E}_\mathrm{eq}
\end{eqnarray}
with ${\cal E}_\mathrm{eq}$ the binding energy at which resonant relaxation begins to dominate classical
relaxation--roughly, the energy of an orbit with radius $a_\mathrm{eq}$
given by equation~(\ref{Equation:aRReqNR}).
Properties of models having the functional form (\ref{Equation:fcore}) for $f$ have been discussed \citep{Merritt2010,AntoniniMerritt2012}.
The distribution of orbital eccentricities, $N(e)$, approximates a delta function,
$N(e)\sim\delta(1-e)$, for $r\ll a_\mathrm{eq}$, since the only orbits that approach closely to the \sbh\ are very eccentric ones.

Figure~\ref{Figure:rholast} suggests that in the Milky Way, the region of strong depletion
would have a radius $< 0.1$ pc.
As noted above, this is probably too small to explain the observed core; but since the depletion
due to resonant relaxation occurs on a $\sim 10^8$ yr timescale, it is likely to be present whatever
the explanation for the larger observed core unless the ``initial conditions'' were extreme.

In the Milky Way, the brightest stars in this region are the S-stars and the stars in the two stellar disks.
Their presence is not inconsistent with the depletion discussed here since these stars must have formed
very recently: less than $\sim 10^8$ years ago in the case of the S-stars, and less than
$\sim 10^7$ years ago in the case of the stellar disks \citep{Schoedel2011}.

The evolutionary models presented here are more relevant to old stellar populations.
Stellar-mass black holes, with masses $\sim 3\msun-30\msun$, are likely to dominate
the mass density inside $\sim 10^{-2}$ pc from the Milky Way \sbh\ \citep{Freitag2006,HopmanAlexander2006L},
due to mass segregation and due to tidal destruction of normal stars.
These are the objects that could become EMRIs, or extreme-mass-ratio inspirals
\citep{SigurdssonRees1997}, of great interest to experimental physicists hoping to detect
low-frequency gravitational waves.
The depletion in $f$ discussed here is likely to have important consequences for the 
steady-state rate of EMRI production.
Exactly what those consequences are can not be stated with certainty yet, 
since the very eccentric orbits
that lead to EMRIs probably evolve in a way that is not well described by the low-$L$ forms
of the diffusion coefficients assumed here.
Rather, these orbits are subject to ``anomalous relaxation,''
the qualitatively different way in which orbits evolve when their precession rate
(due to GR) is much higher than that of the field stars \citep{MAMW2011,AntoniniMerritt2013}.

\section{Summary}
\label{Section:Summary}
Integrations of the Fokker-Planck equation describing $f(E,L,t)$, the phase-space density
of stars around a supermassive black hole (\sbh) at the center of a galaxy, were carried out using
a numerical algorithm described in an earlier paper \citep{Paper1}.
Diffusion coefficients describing both classical and ``resonant'' relaxation were included.
Both steady-state and time-dependent solutions were found.
The principal results follow.

1. Steady-state solutions, with fixed density far from the \sbh, 
are similar to the classical, isotropic, Bahcall-Wolf solution, i.e.
$f\sim |E|^{1/4}$, $n\sim r^{-7/4}$.
However the enhancement of angular momentum diffusion at large binding energies,
due to resonant relaxation, implies a depletion in $f$ at those energies and a corresponding 
density deficit, or ``core.''
The core radius scales approximately with the gravitational influence radius of the \sbh\ and is
 a few percent of that radius. The density within the core is $n\sim r^{-1/2}$.

2. Although the inclusion of resonant relaxation has a substantial effect on the density profile
near the \sbh, the consequences for the \sbh\ feeding rate are much less extreme, since most
stars are scattered into the \sbh\ from orbits that lie outside the resonant relaxation regime and
since $f$ is strongly depleted in that region.
A simple analytic formula is derived, based on the classical diffusion coefficients, that reproduces
the numerically-computed loss rates with good accuracy.
The depletion in $f$ at large binding energies does significantly affect the distribution of orbital elements of
captured stars, in the sense of reducing the contribution from stars on orbits with periapsides
$r_p\approx r_\mathrm{lc}$, the radius of the physical loss sphere.

3. Since energy relaxation times at the centers of galaxies are often very long, time-dependent solutions
were also computed.
Initial conditions were based on a model in which the \sbh\ was preceded by a massive binary.
These models evolve on two timescales: a short timescale during which the core evacuated by the
massive binary is refilled via angular-momentum diffusion; 
and a longer timescale during which diffusion in energy causes the radial distribution of stars 
to approach the Bahcall-Wolf form far from the \sbh.

4. Steady-state cores produced by the effects of resonant relaxation in these models are probably
too small to explain the core observed in the distribution of late-type stars at the center of the Milky Way.
A possible resolution of the apparent discrepancy would be to assume that the nucleus of the Milky
Way has not yet reached a steady state under the influence of gravitational encounters.

5. The depletion in $f$ at large binding energies could have important consequences for the production
of EMRIs, or extreme-mass-ratio inspirals.
A final decision on this question must await a more careful treatment that includes the effects of
``anomalous relaxation,'' the qualitatively different way in which eccentric orbits evolve in the
post-Newtonian regime.

\acknowledgements

This work was supported by the National Science Foundation under grant no. AST 1211602 
and by the National Aeronautics and Space Administration under grant no. NNX13AG92G.

\appendix
\section{Distribution of periapsides}

In the Cohn-Kulsrud boundary layer treatment, at the moment of capture by the \sbh, stars can be 
on orbits with periapsides in the range $0\le r_p \le r_\mathrm{lc}$.\footnote{\bf The calculation described
here is non-relativistic and no special consideration is given to stars that travel near or inside the
hole's event horizon, $r\lesssim r_g$.}
The distribution of orbital integrals (${\cal E}, {\cal R}$) of captured stars is given
by equation (6.57) of \citet{DEGN} (hereinafter DEGN).
Here we recast that equation in terms of orbital elements, including the periapsis distance $r_p$,
and use the result to compute the periapsis distribution for stars in a classical Bahcall-Wolf cusp.
The same algorithm was used to compute the periapsis distribution for the 
Fokker-Planck models in \S~\ref{Section:ResultsSS} (Figure~\ref{Figure:Frarp}).

Differentiating equation (6.57) of DEGN with respect to ${\cal R}$ 
yields the  contribution to the loss cone flux from stars in the energy interval 
${\cal E}$ to ${\cal E} + d{\cal E}$ and angular momentum interval
${\cal R}$ to ${\cal R} + d{\cal R}$:
\begin{eqnarray}\label{Equation:d2NdEdR}
\frac{d^2 F}{d{\cal E} d{\cal R}} &=& 4\pi^2 L_c^2({\cal E}) f({\cal E}, {\cal R}, \tau=1) ,
\ \ \ \ 
{\cal R} \le {\cal R}_\mathrm{lc}({\cal E}) .
\end{eqnarray}
In equation (\ref{Equation:d2NdEdR}), the phase space density $f$ is understood to be
a function of position along an orbit, for any orbit having periapsis inside $r_\mathrm{lc}$ \citep{CohnKulsrud1978}.
The dimensionless variable $\tau$ measures position along such an orbit;
as $\tau$ varies from $0$ to $1$, $r$ increases from periapsis ($r=r_p$, $f=0$),
to apoapsis ($r=r_a$) and back to $r_p$ again (where $f$ has its maximum value along the orbit).
The quantity $f({\cal E}, {\cal R}, \tau=1)$ is expressible in series form as
\begin{eqnarray}
\label{Equation:fatone}
f({\cal E}, {\cal R}, 1) &=& f({\cal E}, {\cal R}_\mathrm{lc})
\left [ 1 - \frac{2}{\sqrt{y_\mathrm{lc}}} \sum_{m=1}^\infty \frac{e^{-\beta_m^2/4}}{\beta_m}
\frac{J_0(\beta_m \sqrt{y})}{J_1(\beta_m \sqrt{y_\mathrm{lc}}}
 \right] 
\end{eqnarray}
(DEGN, equation 6.54).
In equation (\ref{Equation:fatone}), 
$y$ is a dimensionless angular momentum variable defined as 
\beq
y \equiv \frac{{\cal R}}{P({\cal E}) {\cal D}({\cal E})} , \ \ \ \ 
y_\mathrm{lc} \equiv \frac{{\cal R}_\mathrm{lc}}{P({\cal E}) {\cal D}({\cal E})} ;
\eeq
$J_0$ and $J_1$ are Bessel functions of the first kind, and the $\beta_m$ yield
successive zeros of the equation
\beq
J_0\left(\beta\sqrt{y_\mathrm{lc}}\right) = 0.
\eeq
In terms of $\alpha_m \equiv \beta_m \sqrt{y_\mathrm{lc}}$ and
$x \equiv \sqrt{y\; q_\mathrm{lc}}$, equation (\ref{Equation:fatone}) is
\begin{subequations}
\begin{eqnarray}
f({\cal E}, {\cal R}, 1) &=& f_\mathrm{lc}({\cal E}) W(q_\mathrm{lc}, x) , \\
W(q_\mathrm{lc}, x)  &=& 1 - 2\sum_{m=1}^\infty  \frac{e^{-\alpha_m^2 q_\mathrm{lc}/4}}{\alpha_m} \frac{J_0(\alpha_m x)}{J_1(\alpha_m)}
\end{eqnarray}
\end{subequations}
where $q_\mathrm{lc}=q_\mathrm{lc}({\cal E}) = y_\mathrm{lc}^{-1}({\cal E})$ is defined as in equation (\ref{Equation:Defineqlc}) and $f_\mathrm{lc}({\cal E})\equiv  f[{\cal E}, {\cal R}_\mathrm{lc}({\cal E})]$.

In the Kepler potential assumed here, 
$L_c^2({\cal E}) = G^2\mh^2/(2{\cal E})$ so that
\begin{eqnarray}\label{Equation:d2NdEdR2}
\frac{d^2 F}{d{\cal E} d{\cal R}} &=& 2\pi^2 (G\mh)^2 {\cal E}^{-1} f_\mathrm{lc}({\cal E}) W(q_\mathrm{lc},x).
\end{eqnarray}
This can be converted into a distribution in ($r_a,r_p$) using
\begin{eqnarray}
r_a &=& a\left(1+e\right) = \frac{G\mh}{2{\cal E}} \left(1+\sqrt{1-{\cal R}}\right),\ \ \ \ 
r_p = a\left(1-e\right) = \frac{G\mh}{2{\cal E}} \left(1-\sqrt{1-{\cal R}}\right) 
\end{eqnarray}
i.e.
\begin{eqnarray}
{\cal E} &=& \frac{G\mh}{r_a+r_p}, \ \ \ \ 
{\cal R} = \frac{4 r_a r_p}{\left(r_a + r_p\right)^2} .
\end{eqnarray}
The Jacobian is
\beq
\frac{\partial ({\cal E}, {\cal R})}{\partial\left(r_a, r_p\right)} = 
\frac{4{\cal E}^3}{G^2\mh^2} \sqrt{1-{\cal R}} =
4G\mh \frac{r_a-r_p}{\left(r_a+r_p\right)^4} 
\eeq
so that
\begin{eqnarray}\label{Equation:Frarp}
\frac{d^2 F}{dr_a dr_p} = 8\pi^2 G^2 \mh^2 \frac{(r_a-r_p)}{(r_a+r_p)^3} f_\mathrm{lc}({\cal E}) W(q_\mathrm{lc},x) .
\end{eqnarray}
The quantity $x=x({\cal E}, {\cal R}) = \sqrt{{\cal R}/{\cal R}_\mathrm{lc}({\cal E})}$ that appears in
these expressions can be written in terms of the orbital elements as
\begin{eqnarray}
x &=& \sqrt{\frac{r_ar_p}{r_\mathrm{lc}\left(r_a+r_p-r_\mathrm{lc}\right)}}  \;.
\end{eqnarray}
In the limit of nearly-unbound orbits, i.e. ${\cal E}\approx 0$, 
$x\approx \sqrt{r_p/r_\mathrm{lc}} $.

\begin{figure}[h!]
\centering
\includegraphics[angle=-90.,width=7.0in]{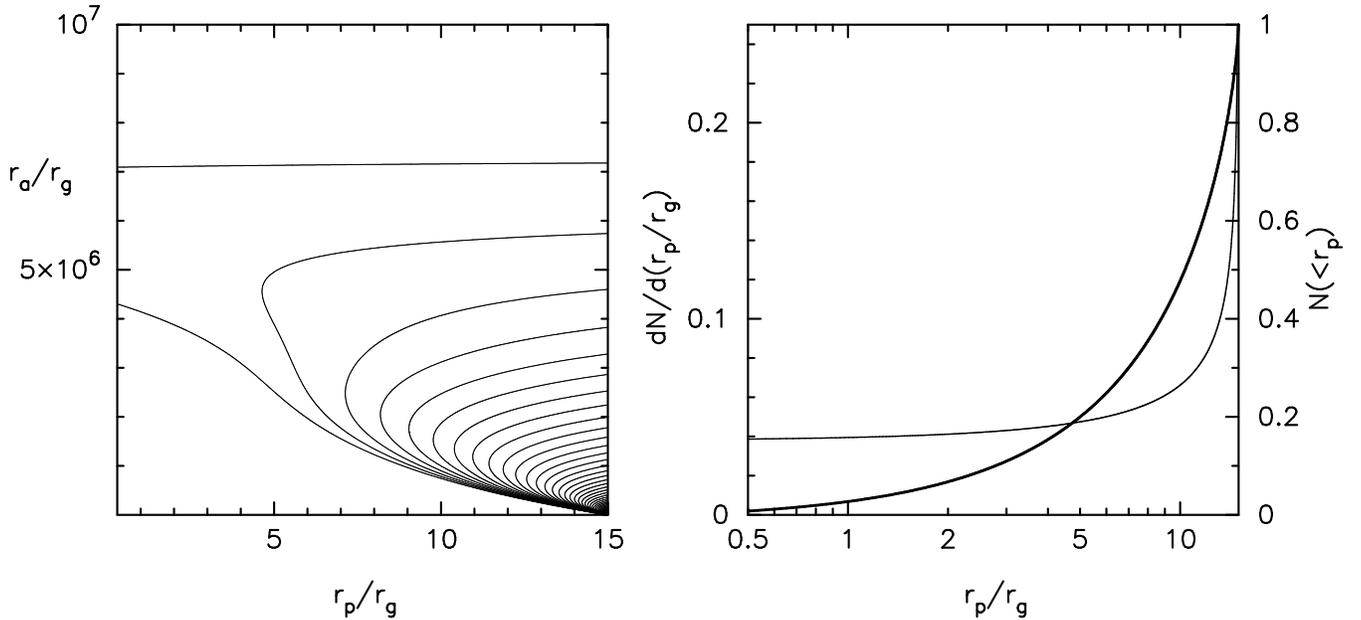}
\caption{Distribution of orbital elements of stars fed to the \sbh\ in a classical Bahcall-Wolf cusp, 
according to the Cohn-Kulsrud boundary-layer solution.
Parameters were $r_m=10^7 r_g$, $m_\star = \mh/(4\times 10^6\msun)$, $\ln\Lambda = 15$,
$r_\mathrm{lc}/r_g=15$.
{\bf Left:} Joint distribution of $r_a$ and $r_p$, equation~(\ref{Equation:Frarp}). 
Contours are spaced uniformly in $\log F$ and
the range in contour values is $10^3$.
{\bf Right:} Thin curve shows $dN/dr_p$, obtained by integrating the function in the left panel 
with respect to $r_a$ at each $r_p$.
Thick curve is the number of stars with periapsides less than $r_p$, obtained by a second
integration with respect to $r_p$. 
Both curves are normalized assuming a unit total number of stars.
\label{Figure:BWrpdist}}
\end{figure}

These relations can be applied to a classical Bahcall-Wolf cusp.
The function $q_\mathrm{lc}({\cal E})$ follows from equations (\ref{Equation:KeplerPeriod}),
(\ref{Equation:Defineqlc}), (\ref{Equation:fofEBW}) and (\ref{Equation:calDNRBWa}):
\beq\label{Equation:qlcBW}
q_\mathrm{lc}({\cal E}) = 8f_0\sqrt{2} \pi^3  \ln\Lambda {\cal R}_\mathrm{lc}^{-1}({\cal E})
\left(G\mh\right)^{1/4} G m_\star r_m^{-5/4}{\cal E}^{-5/4}, \ \ \ \ f_0 \approx 0.036 .
\eeq
We relate $f_\mathrm{lc}({\cal E})$ to $\overline{f}({\cal E})$ using equations (6.61) 
and (6.63) of DEGN:
\beq\label{Equation:flcBW}
f_\mathrm{lc}({\cal E}) \approx \frac{\overline{f}({\cal E})}{1 + q_\mathrm{lc}^{-1} \xi(q_\mathrm{lc}) \ln(1/{\cal R}_\mathrm{lc})}
\eeq
and identify $\overline{f}$ with equation (\ref{Equation:fofEBW}).
The results are shown in Figure~\ref{Figure:BWrpdist}, for a nucleus with
$r_m=10^7 r_g$, $m_\star = \mh/(4\times 10^6\msun)$, $\ln\Lambda = 15$,
and $r_\mathrm{lc}/r_g=15$.
The computation included 5000 terms in the Bessel series and used a $(500 \times 500)$ grid
in $(r_a,r_p)$.
The distribution with respect to $r_p$ can be seen to become nearly uniform for large $r_a$ (full loss cone),
while for small $r_a$ (empty loss cone), the distribution is very strongly peaked toward $r_p=r_\mathrm{lc}$.
Roughly $1/2$ of all captured stars have periapsides in the range
$10r_g \le r_p < 15r_g (= r_\mathrm{lc})$ (right panel).

\clearpage

\end{document}